\documentclass[12pt]{iopart}
\usepackage{iopams}
\usepackage{graphicx}  

\usepackage{subfigure}
\usepackage{color}

\begin{document}

\title[The Ultimatum Game in Complex Networks]{The Ultimatum Game in
  Complex Networks}

\author{R. Sinatra$^{1,2}$, J. Iranzo$^{3,4}$,
  J. G\'omez-Garde\~nes$^{1,4,5}$, L.M. Flor\'{\i}a$^{4,6}$,
  V. Latora$^{1,2}$, and Y. Moreno$^{4,7}$}

\address{$^1$Laboratorio sui Sistemi Complessi, Scuola Superiore di Catania - Via San Nullo 5/i, 95123 Catania, Italy\\
$^2$Dipartimento di Fisica e Astronomia, Universit\`a di Catania, and INFN, Via S. Sofia 64, 95123 Catania, Italy\\
$^3$Centro de Astrobiolog\'ia (CSIC-INTA), 28850 Torrej\'on de Ardoz, Madrid, Spain\\
$^4$Institute for Biocomputation and Physics of Complex Systems (BIFI), University of Zaragoza, 50009 Zaragoza, Spain\\
$^5$ Departmento de Matem\'atica Aplicada, ESCET, Universidad Rey Juan Carlos, 28933 M\'ostoles (Madrid), Spain\\
$^6$ Departamento de F\'{\i}sica de la Materia Condensada,
University of Zaragoza, Zaragoza E-50009, Spain\\
$^7$ Department of Theoretical Physics, University of Zaragoza, 50009 Zaragoza, Spain}

\ead{$^1$roberta.sinatra@ct.infn.it}
\ead{$^5$jesus.gomez.gardenes@urjc.es}

\begin{abstract}
We address the problem of how cooperative (altruistic-like) behavior
arises in natural and social systems by analyzing an ultimatum game in
complex networks. Specifically, three types of players are considered:
(a) empathetic, whose aspiration level and offer are equal, (b)
pragmatic, who do not distinguish between the different roles and aim
to obtain the same benefit, and (c) agents whose aspiration level and
offer are independent. We analyze the asymptotic behavior of pure
populations on different topologies using two kinds of strategic
update rules. Natural selection, which relies on replicator dynamics,
and Social Penalty, inspired in the Bak-Sneppen dynamics, in which
players are subjected to a social selection rule penalizing not only
the less fitted individuals, but also their first neighbors.  We
discuss the emergence of fairness in the different settings and
network topologies.
\end{abstract}

\vspace{2pc}
\noindent{\it Keywords\/}: Network dynamics, Collective phenomena in economic and social systems
%\maketitle

\section{Introduction}
Human cooperation has been the focus of intense debate within the
theoretical framework of evolutionary theories since long time ago
\cite{g02,v04}. In particular, altruistic behavior, in which
individuals perform costly acts for themselves to confer benefits to
the rest of the population, has often been identified as a key
mechanism for cooperation.  A number of theoretical approaches have
been developed to explain the emergence of human altruism. Kin
selection theory \cite{h64} accounts for situations in which it pays
off (inclusive fitness) to help relatives that share some fraction of
the genetic pool. In the absence of such kin relationships, repeated
interactions have also been shown to lead to cooperation, as well as
different kinds of reciprocity mechanisms \cite{v04,ns98,msk02,n06}.
Recently, a series of behavioral experiments in which interactions are
anonymous and one-shot have shown that humans can punish
non-cooperators (altruistic punishment) and reward those individuals
who cooperate (altruistic rewarding)
\cite{v04,fg02,gbbf03,ff03,f05}. This so-called strong reciprocity can
actually explain the observed cooperative behavior in terms of group
and cultural selection. However, standard evolutionary game theory is
still far from explaining how cooperation may arise from selection at
the individual level. Recent steps in this direction \cite{sc05} have
contributed to fill this gap, although a general theoretical framework
is still needed.

On the other hand, recent discoveries on the architecture of
biological, technological and social systems have shown that the
structure of these systems has important consequences on their
dynamical behavior \cite{n03,blmch06}.  In particular, the dynamical
features observed in heterogeneous, scale-free networks, are radically
different from those in homogenous networks.  This difference is due
to the presence of highly connected nodes.  For instance, in epidemic
spreading, the hubs are very efficient in propagating the disease
\cite{pv00,mpv02}, up to the point that in heterogeneous networks the
epidemic threshold vanishes in the limit of infinite system size. In
some other processes, the hubs play the opposite role.  An example is
rumor spreading \cite{mnp04}, where a larger number of "infected"
nodes is obtained in homogeneous networks. Finally, there are
situations where hubs play a more subtle role.  This is the case of
synchronization phenomena \cite{adkmz08}. In many systems, scale-free
(SF) networks exhibit a smaller threshold for the onset of
synchronization. Nonetheless, the stability of the fully synchronized
state is less robust in SF networks than in random graphs.

Motivated by the aforementioned results, studies of evolutionary game
theory models on hetereogenous networks have attracted much attention
in the last years \cite{n06,sf07,sp05,spl06,gcfm07,njponc,jtb,hj08}.
Issues such as the influence of the social structure in cooperative
behavior, as well as the role of the highly connected nodes have been
mainly explored in the context of the Prisoner's Dilemma
\cite{sp05,spl06,gcfm07,jtb}. The results obtained point out that SF
networks are best suited to support cooperation and that hubs play a
fundamental role in spreading cooperation through a positive feedback
mechanism, even when it is expensive. The same kind of results have
been recently reported for public good games \cite{ssp08}.

Here we focus on the Ultimatum Game (UG), another kind of game
extensively used to model altruistic behavior \cite{semUG}, but not
adequately explored in the context of complex networks, though spatial effects have been considered to some extent (see for instance \cite{pns00} for the UG model on regular 1D and 2D lattices). The standard
UG considers that two players bargain to divide a fixed reward between
them. Suppose that one of these players acts as proposer offering a
division of the reward. The other, henceforth called respondent, can
accept or reject this proposal, but cannot counteroffer. If the
respondent accepts, the reward is divided as agreed, otherwise both
receive nothing. For a one-shot game played anonymously, the rational
solution (subgame perfect Nash equilibrium solution) is that in which
the proposer would offer the smallest possible share and the
respondent would accept it. However, plenty of experimental results
point out that the rational solution is not what actually happens. For
instance in the social context, it has been shown that the mean offer
is usually between $40\%$ and $50\%$ and that offers below $20\%$ of
the reward are often rejected \cite{book,henrich}. This has been interpreted
as an example of altruistic punishment \cite{fg02,f05}, i.e., the
tendence to impose sanctions on unfair individuals with a cost for the
punisher. However, costly punishment has been proven \cite{drfn08} to
be maladaptive (winners do not punish) which leaves open the question
on how this trait has evolved.
%In contrast to other works \cite{ej04,hj08}, 

We implement here two kinds of evolution rules (see below): one is
fitness-dependent and is based on a pairwise comparison, in the spirit
of \cite{gcfm07,jtb}, and the second one is inspired in the
Bak-Sneppen model \cite{bs,morenosoc,ej04} of punctuated equilibrium.
Summing up, in the present work, we study an UG model on
Erd\"os-R\'enyi and Scale-free networks with three different kinds of
settings of the parameters characterizing the players. The asymptotic
evolutionary states reached following the two update rules cited above
are analyzed and compared in the three different frameworks.

% where: {\em i)} players are placed on the nodes of a graph and play with their
% neighbors; {\em ii)} a selection rule in which the less fitted
% individuals are replaced together with their neighbors is implemented.

% Our results point out that the interaction topology has a strong
% influence in the emergence of altruism. In particular, in SF networks,
% highly connected players show a striking generous behavior, which has
% a feedback effect over the macroscopic level of altruism observed in
% the population, since it allows most of the players to adopt selfish
% behaviors in order to increase their benefits. As a result, there is a major
% diversity of strategies at the microscopic level: agents do not play
% all in the same way, but base their choices on the number of
% neighbors, or, equivalently, on the number of possible rewards to
% divide.

\section{The model}

In our model we consider $N$ individuals associated to the nodes of a
graph. The graph topologies we will study are of two different kinds:
Erd\"os-R\'enyi (ER) and Scale-free (SF) networks. An ER network is
characterized by a degree distribution that decays exponentially fast
for large $k$, while in a SF network the degree distribution follows a
power-law of the form $P_k\sim k^{-\gamma}$. We consider SF networks
with $\gamma\approx 3$ \cite{gm06}.  Therefore, while in ER networks
the number of contacts shared by individuals shows a finite variance,
in SF networks we find nodes, usually referred to as hubs, that
interact with a large fraction of the population.

\subsection{Playing the Ultimatum Game}

The individuals on the nodes of the aforementioned networks play the
Ultimatum Game (UG). At each time step, each individual plays a round
robin of the game with all his neighbors, as dictated by the graph. In
each round, individuals play the UG twice with each neighbor, both as
proposers and as respondents.  The reward to divide in each of these two
games is equal to $1$. An individual $i$ ($i=1,...,N$) is
characterized by two parameters: $p_{i}$, $q_{i}$ $\in [ 0,1]$. When
$i$ acts as proposer it offers a division $p_i$ of the reward, so that
the respondent will earn $p_i$ if the proposal is accepted. Instead,
when agent $i$ plays as respondent, it will accept only offers larger
than its acceptance threshold $q_i$.
Therefore, when two individuals ($i,j$) bargain, their payoffs,
$\Pi_{i}$ and $\Pi_{j}$, evolve according to the following rules:
\begin{itemize}

\item{Player $i$ offers the amount $p_i$ to $j$. If $p_i \ge q_j$, the
  offer is accepted and the payoff of $i$ and $j$ are incremented by
  $\Delta\Pi_{ij}^{O}=(1 - p_i)$ and $\Delta\Pi^{R}_{ji}=p_i$
  respectively. Conversely, if $p_i < q_j$, agreement is not possible
  and both players get nothing and their payoffs remain the same,
  $\Delta\Pi_{ij}^{O}=\Delta\Pi^{R}_{ji}=0$. }

\item{When player $i$ is the respondent, the same rules
  apply. Therefore, upon agreement ($p_j > q_i$), players $i$ and $j$
  increase their payoffs by $\Delta\Pi^{R}_{ij}=p_j$ and
  $\Delta\Pi^{O}_{ji}=(1-p_j)$ respectively.}

\end{itemize}
The final payoffs of a node $i$ after playing with all its neighbors
is
\begin{equation}
\Pi_i=\sum_{l \in  \Gamma_i}(\Delta\Pi_{il}^{O}+\Delta\Pi_{il}^{R})\;,
\end{equation}
 where $\Gamma_i$ denotes the set of $i$'s neighbors.

In the following, we will study three different settings for the
values of the parameters $p_i$ and $q_i$:
\begin{enumerate}
 \item[(A)] For each agent $i$, $p_i=q_i$ \cite{pn02}. This is usually called a
   fair or empathetic setting since each agent offers the
   same reward it is disposed to accept;
\item[(B)] For each agent $i$, $p_i=1-q_i$ \cite{nps00}. This is a
  role-ignoring or pragmatic setting since each agent wants to
  get the same reward both as respondent and as proposer;
\item[(C)] The values of $p_i$ and $q_i$ are independent for each agent.
\end{enumerate}

The second choice B stands for a situation in which players do not
differentiate between roles (\emph{role-ignoring} agents). In other
words, regardless of whether they act as proposers or responders, they
are determined to obtain a fixed quantity from each interaction, so
that $q_i=1 - p_i$ \cite{nps00}. This situation is in contrast with
the case of an empathetic or fair and {\em role-distinguishing}
setting A, according to which individuals do distinguish among
roles. In this case the threshold of acceptance is set equal to the
one for proposals ($q_i=p_i$), so as to get half of the total stake on
average. Finally, in the third setting C the quantity offered and the
threshold of acceptance are completely independent as in the original
formulation of the UG.

Note, that in both cases A and B, the corresponding relations
$p(q)$ allow to obtain simple rules for the conclusion of a deal
between two players. Given that the offer $p_i$ proposed by player $i$
is accepted by $j$ only if $p_i\ge q_j$ we have the two following
scenarios:
\begin{enumerate}
\item[(A)] Case $p=q$: if $p_{i}\neq p_j$ $i$ and $j$ always
  conclude a deal, but only in one of the two directions. In
  particular, the accepted offer is the largest one: $\max\{p_i,p_j\}$.
  If for example $p_i>p_j$, the payoffs are incremented by:
\begin{eqnarray}
  \Delta\Pi_{ij}&=&\Delta\Pi^{O}_{ij}=1-p_i\;,
\label{eq:payoffA1}
  \\ 
  \Delta\Pi_{ji}&=&\Delta\Pi^{R}_{ji}=p_i\;.
\label{eq:payoffA2} 
\end{eqnarray}
If $p_i=p_j$ the deal is concluded in both directions and their
payoffs are incremented in $\Delta\Pi_{ij}=\Delta\Pi_{ji}=1$, which is
the maximum possible reward after the interaction between two players
of type A.

\item[(B)] Case $p=1-q$: both players $i$ and $j$ will obtain
  reward both as proposers and respondents if the condition
  $p_i+p_j\ge 1$ is verified. In this case, their payoffs are
  incremented by
  \begin{eqnarray}
  \Delta\Pi_{ij}&=&\Delta\Pi^{O}_{ij}+\Delta\Pi^{R}_{ij}=(1-p_i)+p_j\;,
  \label{eq:payoffB1}
  \\
  \Delta\Pi_{ji}&=&\Delta\Pi^{O}_{ji}+\Delta\Pi^{R}_{ji}=(1-p_j)+p_i\;.
  \label{eq:payoffB2}
\end{eqnarray}
When $p_i+p_j< 1$ no payoff is obtained in the round.
\end{enumerate}

\begin{figure}
\begin{center}
\includegraphics[width=1.0\textwidth]{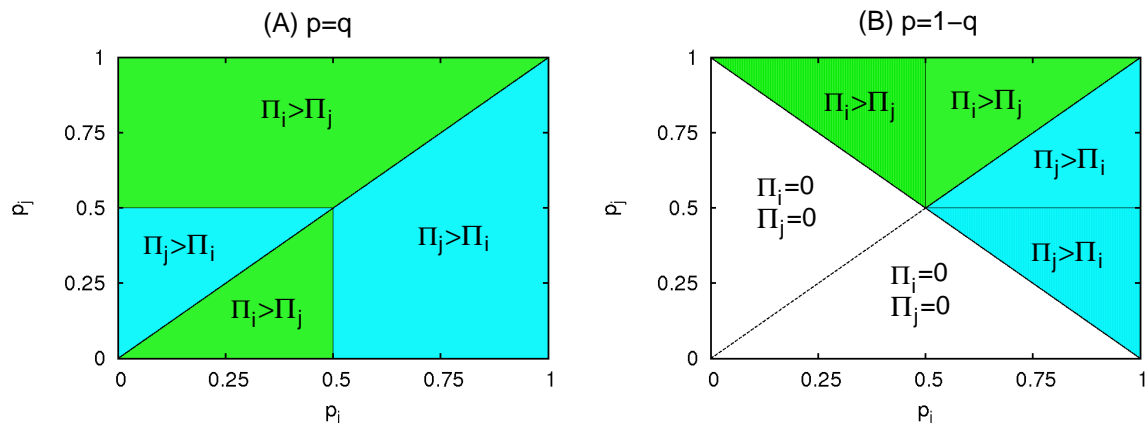}
\end{center}
\caption{The figures show the partition of the strategies space of a two UG
  players ($i$ and $j$) into different regions. Each of these regions
  is labelled according to the payoffs ordering: green areas
  correspond to the case $\Pi_i>\Pi_j$, blue areas to
  $\Pi_j>\Pi_i$. The regions in white correspond to the
  case of a zero reward for both players $\Pi_i=\Pi_j=0$. (See text for
  the details).}
\label{fig:0}
\end{figure}

We illustrate the different ordering in payoffs for the two type of
players in Figure \ref{fig:0}.

\subsection{Updating the strategies}

Once a player has bargained with all its neighbors, the accumulated
payoff drives the update of their strategies. This update process
takes place at the individual level, in the same spirit of
\cite{sc05}, and follows two different schemes:
\begin{itemize}

\item {\em Natural selection}: In this framework, originally introduced
  in \cite{helb92-1,helb92-2}, each player $i$ in the network selects at random
  one neighbor $j$ and compares its payoff $\Pi_i$ with the one of
  $j$, $\Pi_j$. If $\Pi_j>\Pi_i$, player $i$ adopts the strategy of
  $j$, ($p_j$, $q_j$), for the next round of the UG with a probability
  proportional to the payoff difference:
  \begin{equation}
    P_{ij}=\frac{\Pi_j-\Pi_i}{2\max\{k_i,k_j\}}\;.
\label{eq:REP}
  \end{equation}
where $k_i$ and $k_j$ are the degrees of $i$ and $j$ respectively.
Instead, if $\Pi_i\leq \Pi_j$, $i$ keeps his strategy for the
following round.

\item {\em Social penalty}: The player with lowest payoff in the whole
  population together with its neighbors, no matter how
  wealthy they are, are removed. These agents are
  replaced in their nodes by new players with random strategies (so that
  they only inherit their contacts).

 \end{itemize}
In the case of {\em Natural selection}, there is a pairwise comparison
thanks to which fittest strategies are replicated with a rate
proportional to their success, with the result of eventually spreading
over the whole population \cite{gcfm07}. As we will discuss, these
dominant strategies might not promote the welfare of the population
since it acts at a local level. On the contrary, {\em Social penalty}
acts at the global level; the removal of all the neighbors of the
least-fitted agent is a catastrophic effect triggered by his
extinction (see \cite{bs} for a discussion on the evolutionary
justification of this updating rule) and not related to individuals'
fitness but to the network of interactions. This undiscriminating (and
likely unfair) social penalty is imposed on those agents that in the
community are responsible for the low fitness of the dying agent; thus
it is quite different from the current notion of (altruistic)
punishment commented above. With this evolutionary rule, a player, in
order to survive, has to take care not only of his payoff, but also of
the neighbors' one: if an individual exploits its neighborhood so that
it takes a large stake of the total reward, it would risk to be
dropped out of the game as a result of one of its neighbors being that
with the lowest payoff in the population of players.  In both the {\em
  Social penalty} and {\em Natural selection} contexts, after the
implementation of the update rule, the payoffs of the agents are reset
to zero. This means that players have no memory of the previous round
payoffs, although they keep their strategies; consequently it is a
one-shot game and no mechanism of reputation has been explicitly
introduced \cite{nps00}.

In the following, we will analyze the scenarios concerning these two updating rules in
ER and SF topologies for the three strategic settings A, B and C introduced above.

\section{The Ultimatum game with Natural selection}
The behavior of players of type $A$ (empathetic) and $B$ (pragmatic)
can be easily predicted in a well-mixed population when a
replicator-like dynamics is at work. Because of this, in the following
two subsections we will first discuss the evolution of the game in a
well-mixed population and then compare it with the numerical results
obtained for homogeneous and heterogeneous networks.

\subsection{Networks of type A players ($p=q$)}
\label{A.RD}

As mentioned above, in a round robin between two empathetic ($q=p$)
players $i$ and $j$ the largest offer, say $p_i$, is always accepted
by the player offering less, hence $j$, and the payoff obtained will
be those of eqs. (\ref{eq:payoffA1}) and (\ref{eq:payoffA2}). In the
case of $p_i >p_j$, two situations are possible: ({\em i}) $p_i>0.5$,
so that $\Pi_{j}>\Pi_{i}$ and ({\em ii}) $p_i<0.5$, yielding 
$\Pi_{i}>\Pi_{j}$ (see Figure \ref{fig:0}.a).

In the case of the dynamics of a well-mixed population where all the
individuals interact with the rest of the players, given the
distribution $D(p)$ of offers in the population one finds that the
payoff received by strategist offering $x$ is $\Pi(x)=G(x)+\langle
p\rangle - H(x)$ where:
\begin{eqnarray}
G(x)&=&(1-x)\int_{0}^{x}D(p){\mbox d}p\;,
\\
H(x)&=&\int_{0}^{x}pD(p){\mbox d}p\;,
\\
\langle p\rangle&=&\int_{0}^{1}pD(p){\mbox d}p\;.
\end{eqnarray}

In the case of replicator dynamics, the increase or decrease of the
fraction of players using strategy $x$ is determined by
$\Pi(x)-\langle \Pi\rangle$, being $\langle \Pi\rangle$ the average
payoff in the population. For a uniform distribution $D(p)=1$ one
obtains $\Pi(x)-\langle \Pi\rangle=x-3x^2/2$ and one concludes that
from an initial uniform distribution the highest values of $p$ will
soon become extinct, and the highest increase in frequency will occur
for values centered at $x=1/3$. Once most of players use offers below
$1/2$, the selective advantage is for players with higher $p$ (below
$1/2$). Thus, one expects that the values of $p$ will concentrate at
$p=1/2$. This two stage dynamics will be obtained also in the context
of complex networks.

\begin{figure}
 \centering
\subfigure[ER $p=q$]
   {\includegraphics[width=7cm]{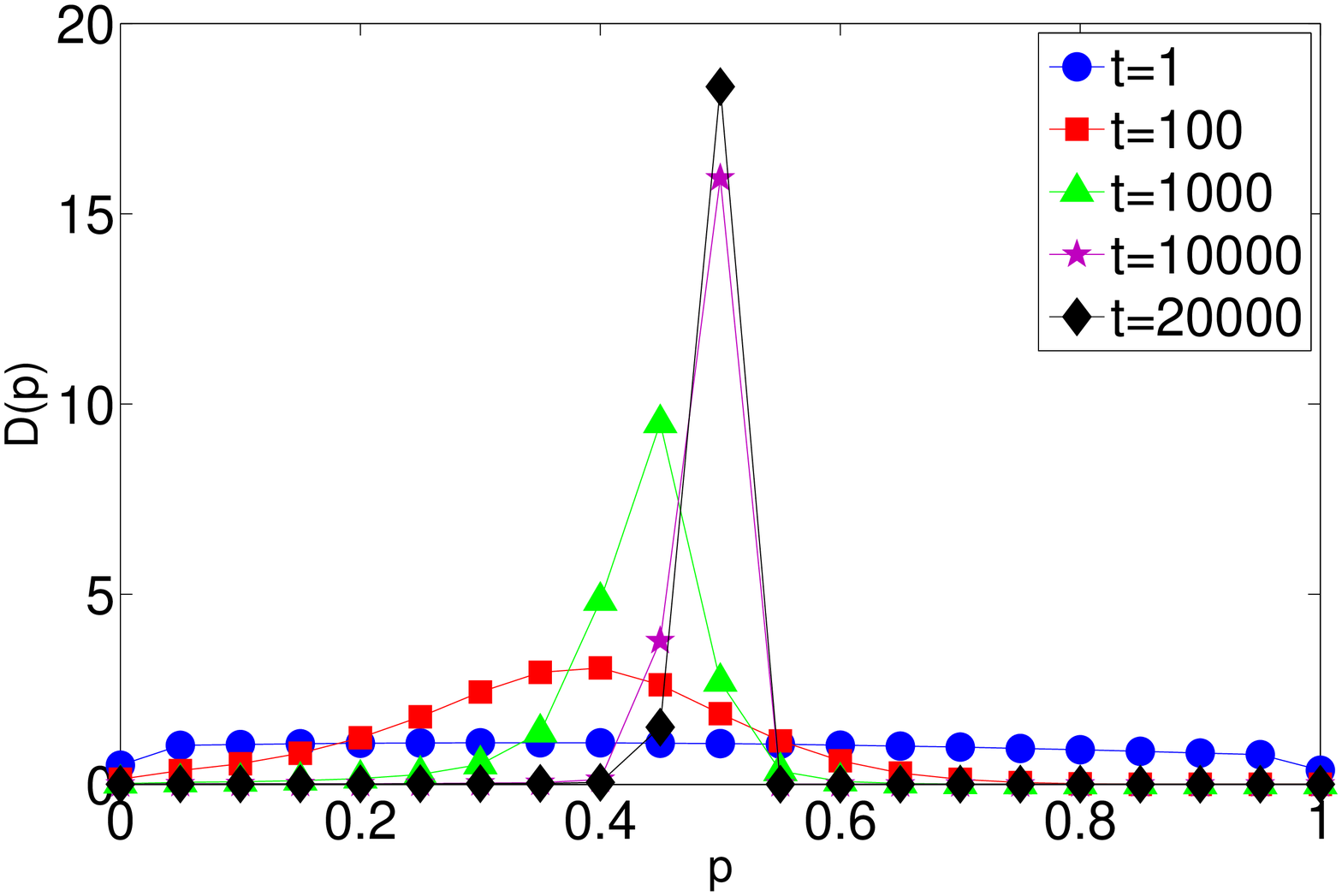}}
\subfigure[SF $p=q$]
   {\includegraphics[width=7cm]{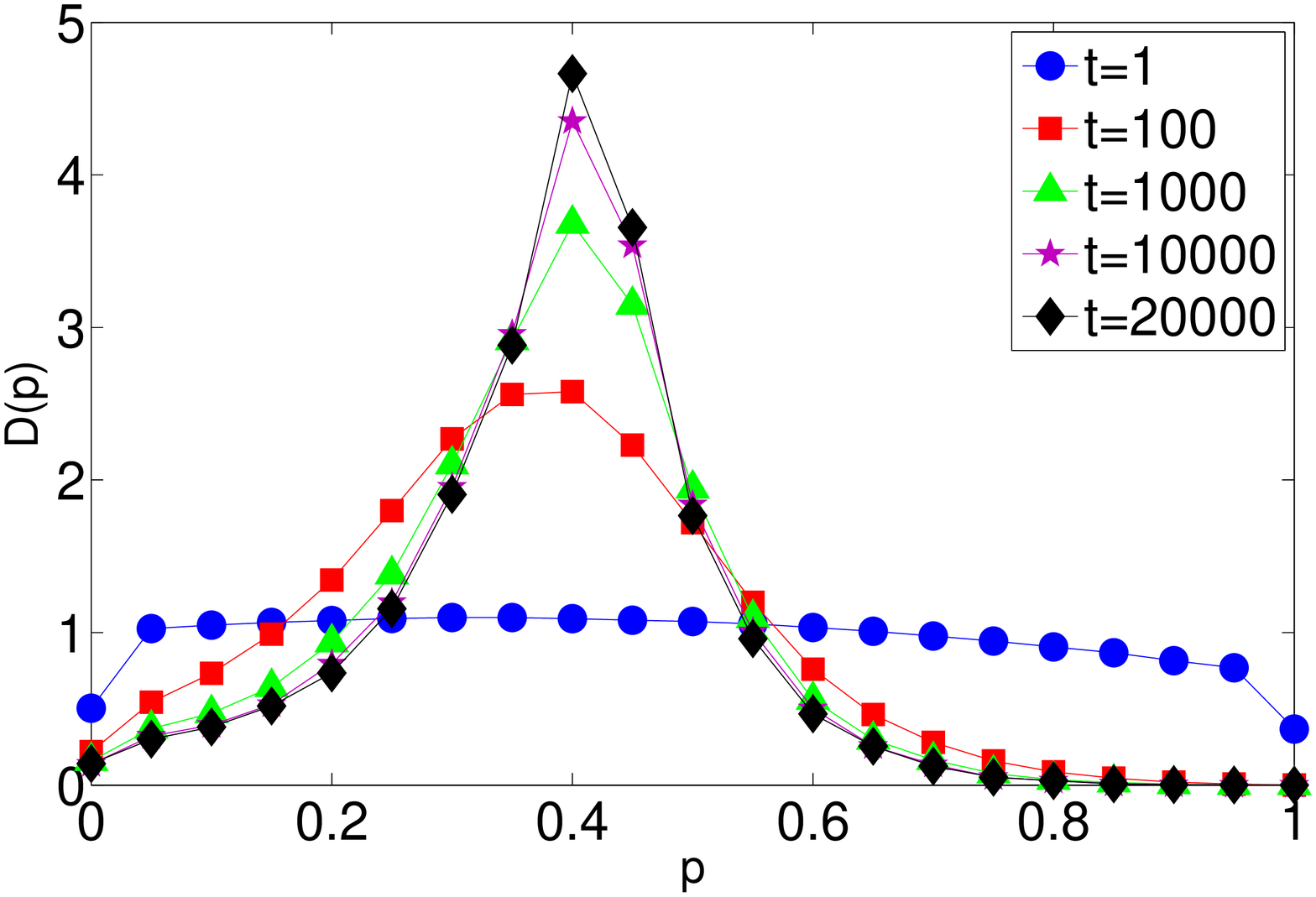}}
\subfigure[ER $p=1-q$]
   {\includegraphics[width=7cm]{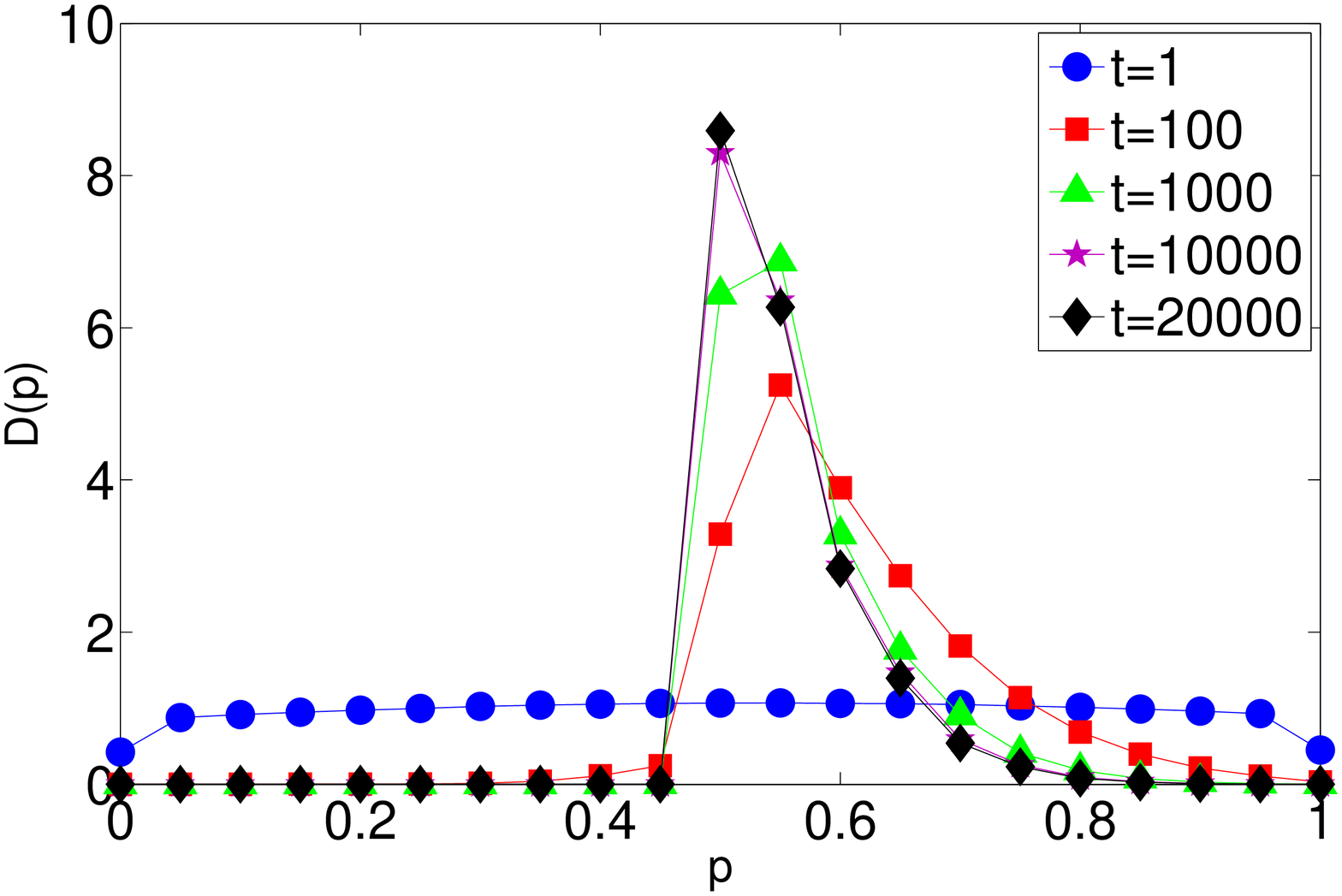}}
\subfigure[SF $p=1-q$]
   {\includegraphics[width=7cm]{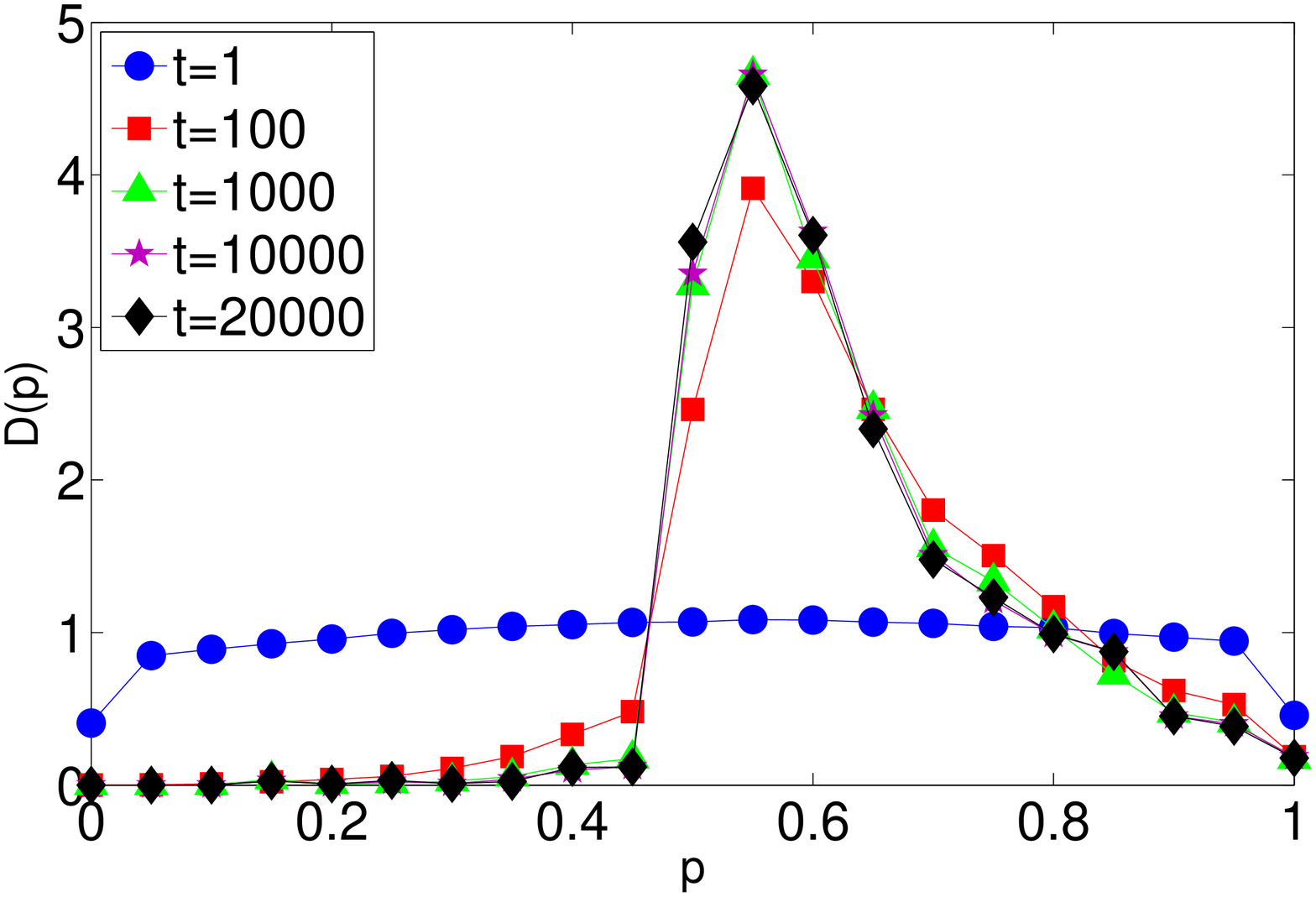}}
\caption{Distribution of offers $D(p)$ for ER and SF networks in the
  cases $p=q$ [(a) and (b)] and $p=1-q$ [(c) and (d)] when a
  Replicator-like update rule, eq. (\ref{eq:REP}), is at work.}
\label{fig:RD1}
 \end{figure}

We show the results obtained with this dynamics on top of ER and SF
networks. In both cases the networks have $N=10^4$ nodes and average
degree $\langle k \rangle=4$. The evolutionary dynamics starts
assigning to each individual of the population a random offer $p_i$
(and thus $p_i=q_i$) uniformly distributed in the interval
$[0,1]$. Then, we follow the system evolution for a number of time
steps until a stationary regime is reached. The results presented are
averaged over at least $10^3$ realizations of both the underlying
network and the initial conditions.

Figures \ref{fig:RD1}.a and \ref{fig:RD1}.b show the time evolution of
the distribution of offers $D(p)$ in the population for both ER and SF
networks. It is evident that for ER networks the distribution $D(p)$
after $t=2\cdot 10^4$ generations shows qualitatively the shape
predicted using the well-mixed assumption. Moreover, the two-stage
evolution explained above is also confirmed by looking at the time
evolution of $D(p)$. From $t=1$ to $t=10^2$ the strategists with
$p>0.5$ are removed and invaded by those players with low values of
$p$. After this initial stage, the flow of strategies goes from low
values of $p$ towards $p=0.5$, reaching the final distribution peaked
at $p\simeq 0.5$ with a fast decaying tail at $p<0.5$.

In the case of SF networks the asymptotic distribution of offers
$D(p)$ becomes broader with respect to ER graphs. Remarkably the
two-stage process is also observed since most of strategies with large
values of $p$ are removed in the first time steps. On the other hand,
at variance with ER networks, some strategies with $p>0.5$ survive in
the final population. This result is the consequence of having
individuals, named hubs, with large degree $k_h>\langle k\rangle$.
The analysis of a ``coarse grained'' picture of %{\em (i)} a
degree-homogeneous population of size $N$ and mean degree $\langle
k\rangle$
%and {\em (ii)} 
with an individual connected to a large number, $k_h$, of individuals
of this population can help us to understand what takes place for SF
networks. Suppose that the population has reached its internal equilibrium and
therefore $p_i\simeq0.5$ for all its members. In the case $p_h<0.5$
(selfish hubs), a hub obtains a payoff $\Pi_h=k_h/2$ while the members
of the population connected to the hub obtains $\Pi_i=\langle
k\rangle+1/2$. In this case, the hub survives ({\em i.e.} satisfies
$\Pi_h\ge \Pi_i$) for every value $p_h<0.5$ provided that $k_h\ge
2\langle k\rangle +1$, a condition that is easily verified in SF for
large degree nodes. On the other hand, if $p_h>0.5$ (generous hubs) we
have $\Pi_h=k_h(1-p_h)$ and on average $\langle \Pi \rangle =\langle
k\rangle + p_h$ for the individuals connected to the hub. Therefore,
if generous hubs are to survive in the system they cannot offer more
than $p_h\leq (1-\langle k\rangle/k_h)$. This maximum offer tend to
$1$ as $k_h$ grows, thus explaining the existence of a tail for
$p>0.5$ in the distribution $D(p)$ of SF networks. In both
  cases, the strategy of hubs is eventually replicated by the rest of the
  population and after enough generations the payoff of the hub is
  $\Pi_h=k_h$ while $\langle \Pi \rangle =\langle k\rangle$ for its
  neighbors. Hence, heterogeneity can help the fixation of altruistic
  behavior in nodes provided they have a large number of contacts to
  obtain enough payoff.

\subsection{Networks of type B players ($p=1-q$)} 
Let's now focus on the case B. In this context, two players $i$ and
$j$ conclude a deal only when $p_i+p_j\geq 1$. If this condition
holds, the consensus is automatically reached in both directions, and
the payoffs of the players are those specified in
eqs. (\ref{eq:payoffB1}) and (\ref{eq:payoffB2}). The line (see Figure \ref{fig:0}.b) $p_i=1-p_j$ delimitates
the area of unsuccessful strategies (below the line), since no payoff
is obtained, from that of the successful ones (above the line).  This
latter region can be further divided into two triangular areas: that
of $p_i>p_j$, yielding $\Pi_j>\Pi_i$, and the one of $p_j>p_i$, giving
$\Pi_i>\Pi_j$. Obviously, the border between the two regions is
specified by $p_i=p_j$ (see Figure \ref{fig:0}.b).

For a well mixed with a distribution density of offers $D(p)$, the
payoff of strategist $x$ is $\Pi(x)=G'(x)+ H'(x)$ where:
\begin{eqnarray}
G'(x)&=&(1-x)\int_{1-x}^{1}D(p){\mbox d}p\;,
\\
H'(x)&=&\int_{1-x}^{1}pD(p){\mbox d}p\;.
\end{eqnarray}
For an initial uniform density $D(p)=1$, one obtains $\Pi(x)-\langle
\Pi\rangle=-3x^2/2+2x-1/2$ whose graph is mirror-symmetric around
$x=1/2$ of the one obtained for type A players. Thus, one expects a
fast extinction of lowest offers and an initially higher growth of
offers around $2/3$. Once offers below $1/2$ become extinct, one
easily realizes that for any arbitrary corresponding density ($D(p)=0$
for $p<1/2$), $\Pi(x)-\langle \Pi\rangle=\langle p\rangle -x$ so that
the selective advantage is for offers as close to $p=1/2$ as possible.
Therefore one expects a progressive displacement to $p=1/2$ values of
the maximum of the evolving density.

When performing simulations of B players on ER networks, similarly to
what happens for A players (see \ref{A.RD}), the asymptotic
distribution of offers agrees with the well-mixed predictions, as
Figure \ref{fig:RD1}.c confirms.  Here the distribution at $t=2\cdot
10^4$ shows a peak at $p=0.5$ and a fast decaying tail for
$p>0.5$. This tail proves that strategies moves towards $p=0.5$ from
the right, {\em i.e.} from the successful region of the
$N$-dimensional space. Remarkably, the strategies with $p<0.5$ are
totally removed from the population already at $t=100$.

The distribution of offers $D(p)$ in SF networks, Figure
\ref{fig:RD1}.d, shows also a peak around $p=0.5$ but with a tail for
$p>0.5$ decreasing slower than in ER networks.  This behavior can be
explained again with the presence of highly connected
players. Following the same argument used for A players, a hub with an
offer $p_h>0.5$, connected to a large number $k_h$ of individuals with
$p\simeq0.5$ and mean degree $\langle k\rangle$, obtains a payoff
$\Pi_h=k_h(3/2-p_h)$ whereas for the individuals connected to the hub
on average $\langle \Pi \rangle=\langle k\rangle+p_h+1/2$. In this
setting, the hub will survive and spread its strategy provided
$p_h\leq (3/2-\langle k\rangle/k_h)$. Therefore offers of hubs can
also reach $p=1$ as observed in the distribution $D(p)$ for SF
networks. Similarly as in the case of type A players, once the
  hub's neighbors have imitated its strategy the payoffs of the hub
  and its neighbors are $\Pi_h=k_h$ while $\langle \Pi \rangle
  =\langle k\rangle$ respectively. In the case $p_h<0.5$, since the
  condition $p_h+p_i\ge 1$ is no longer verified, the region with
  $p<0.5$ keeps on being forbidden, in agreement with the sharp decay
  of $D(p)$ in Figure \ref{fig:RD1}.d.

Interestingly, at variance with the case of type A players in which
the unsuccessful strategies of the well-mixed case ($p>0.5$) are
allowed to the high degree nodes of SF networks, in the case of B
players the unsuccessful region of strategies of the well-mixed limit
($p<0.5$) is always empty, regardless of the underlying topology of
interactions.

\begin{figure}
 \centering
\subfigure[ER $(p,q)$]
   {\includegraphics[width=7cm]{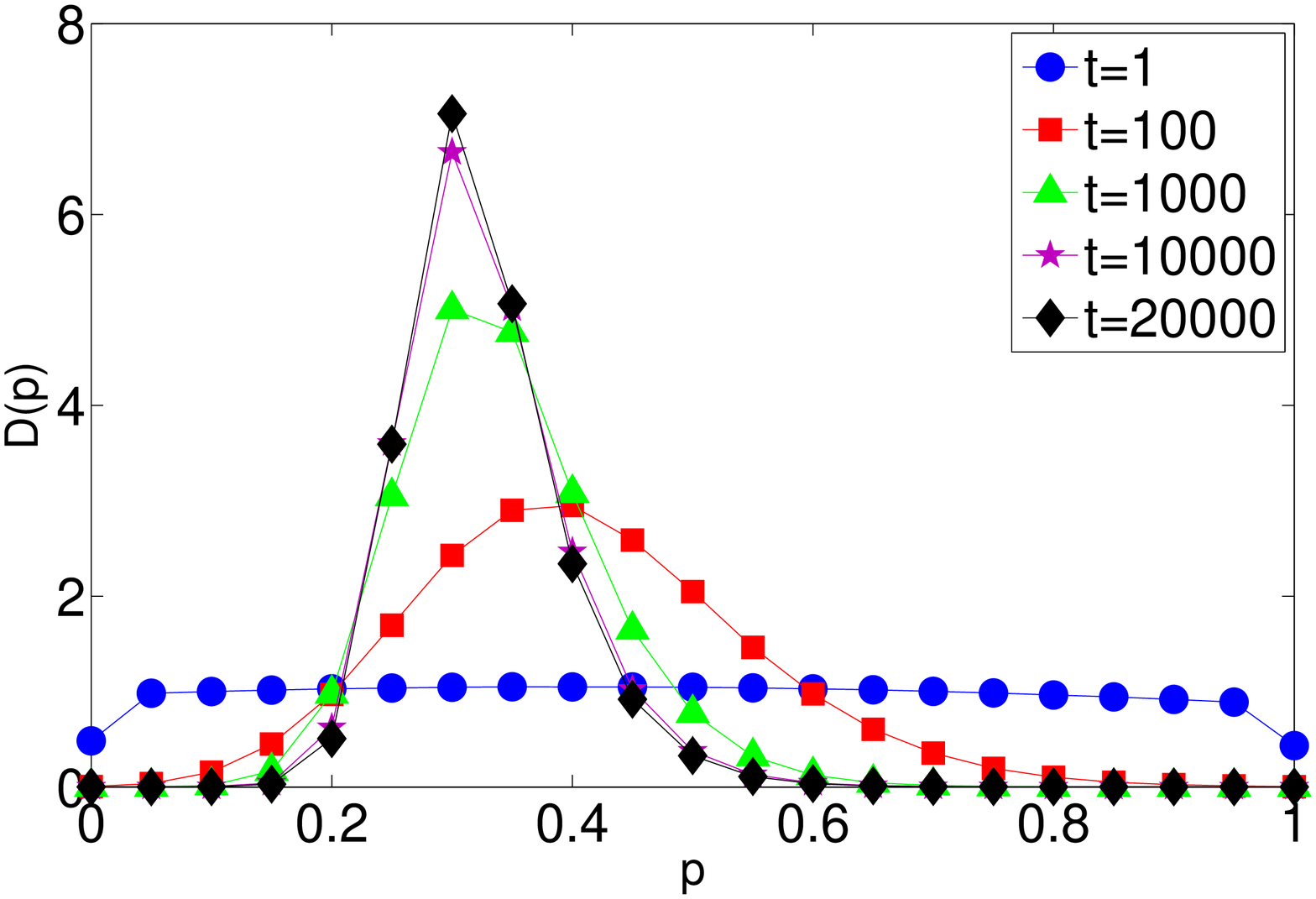}}
\subfigure[SF $(p,q)$]
   {\includegraphics[width=7cm]{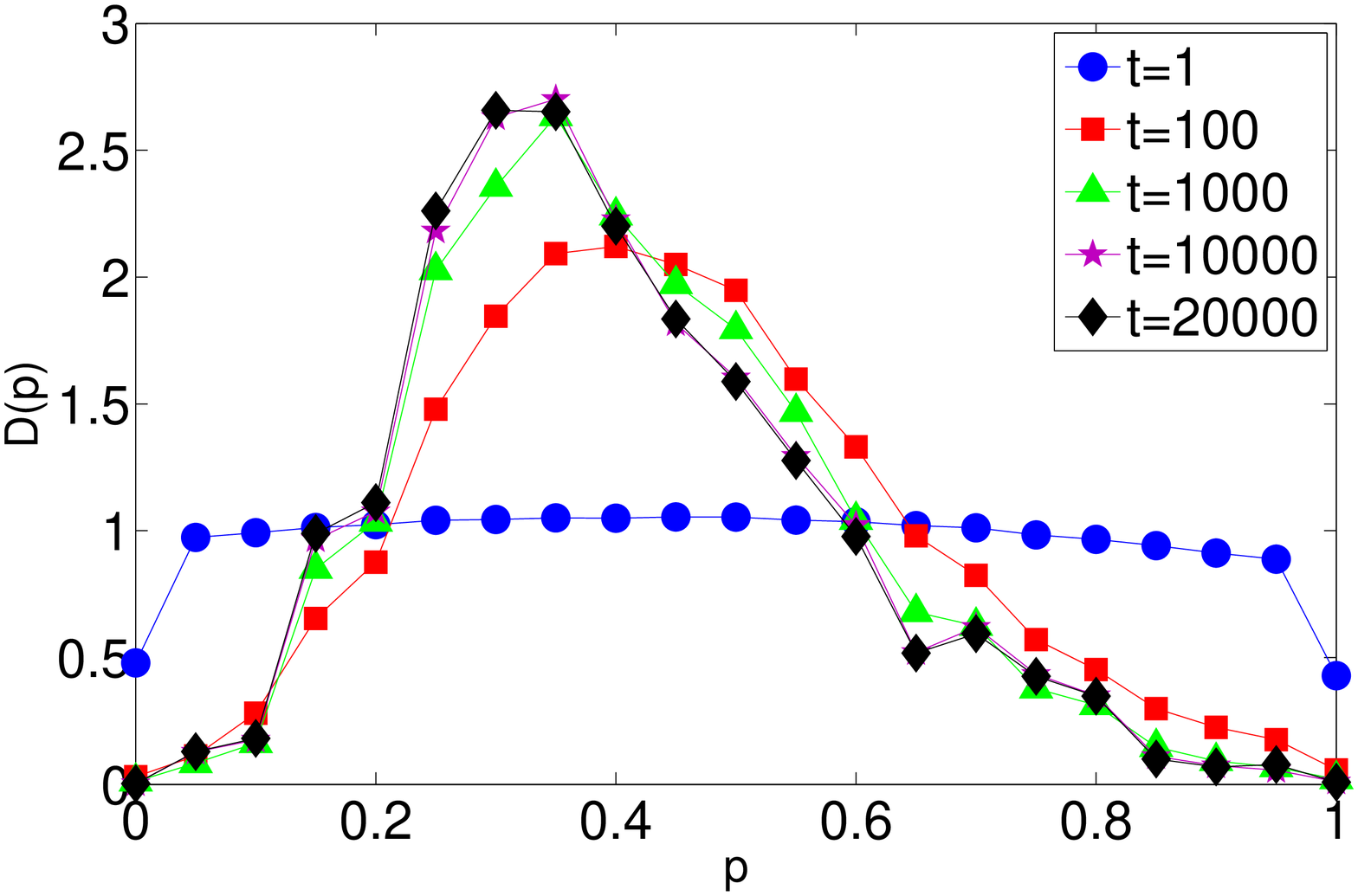}}
\subfigure[ER $(p,q)$]
   {\includegraphics[width=7cm]{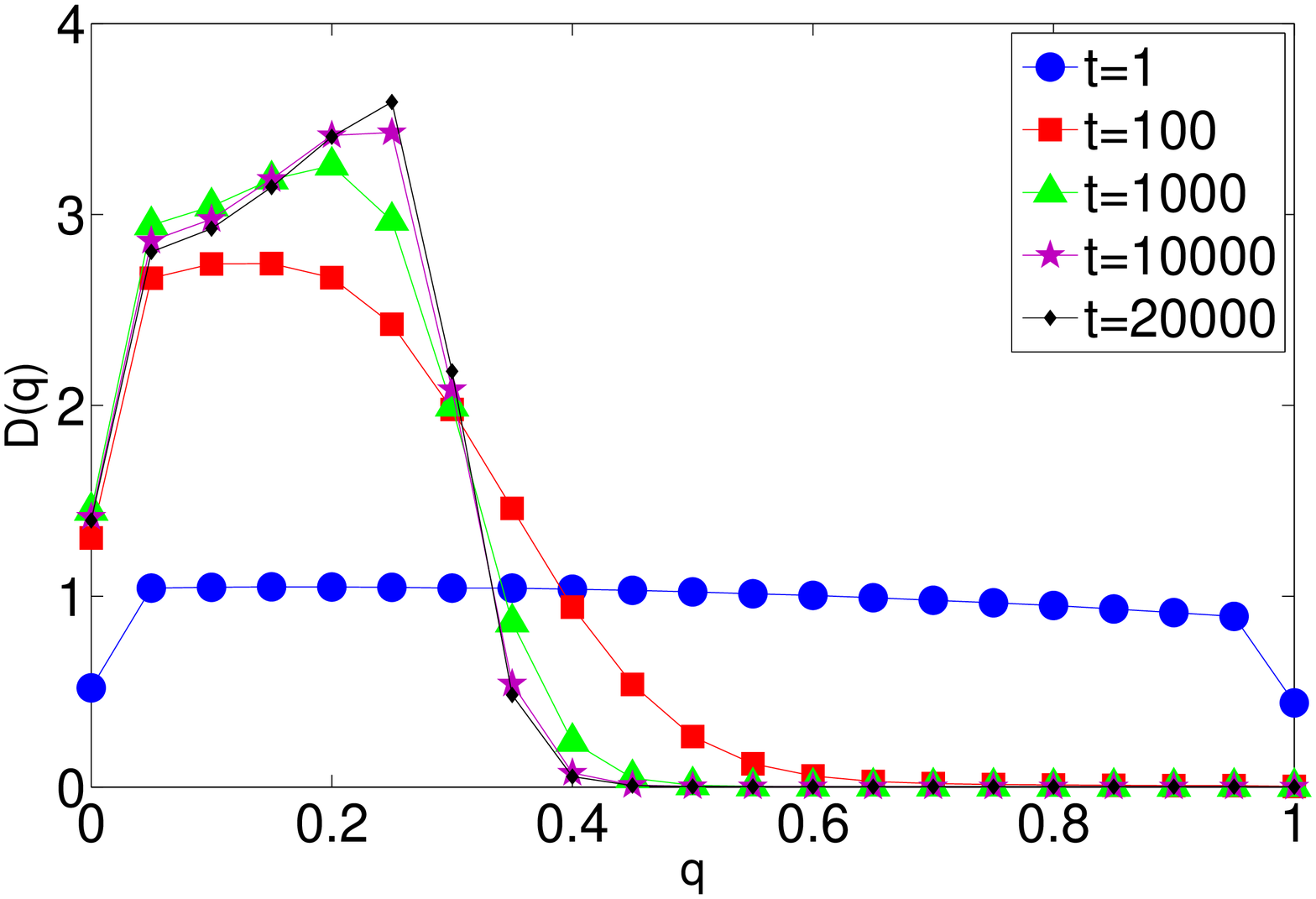}}
\subfigure[SF $(p,q)$]
   {\includegraphics[width=7cm]{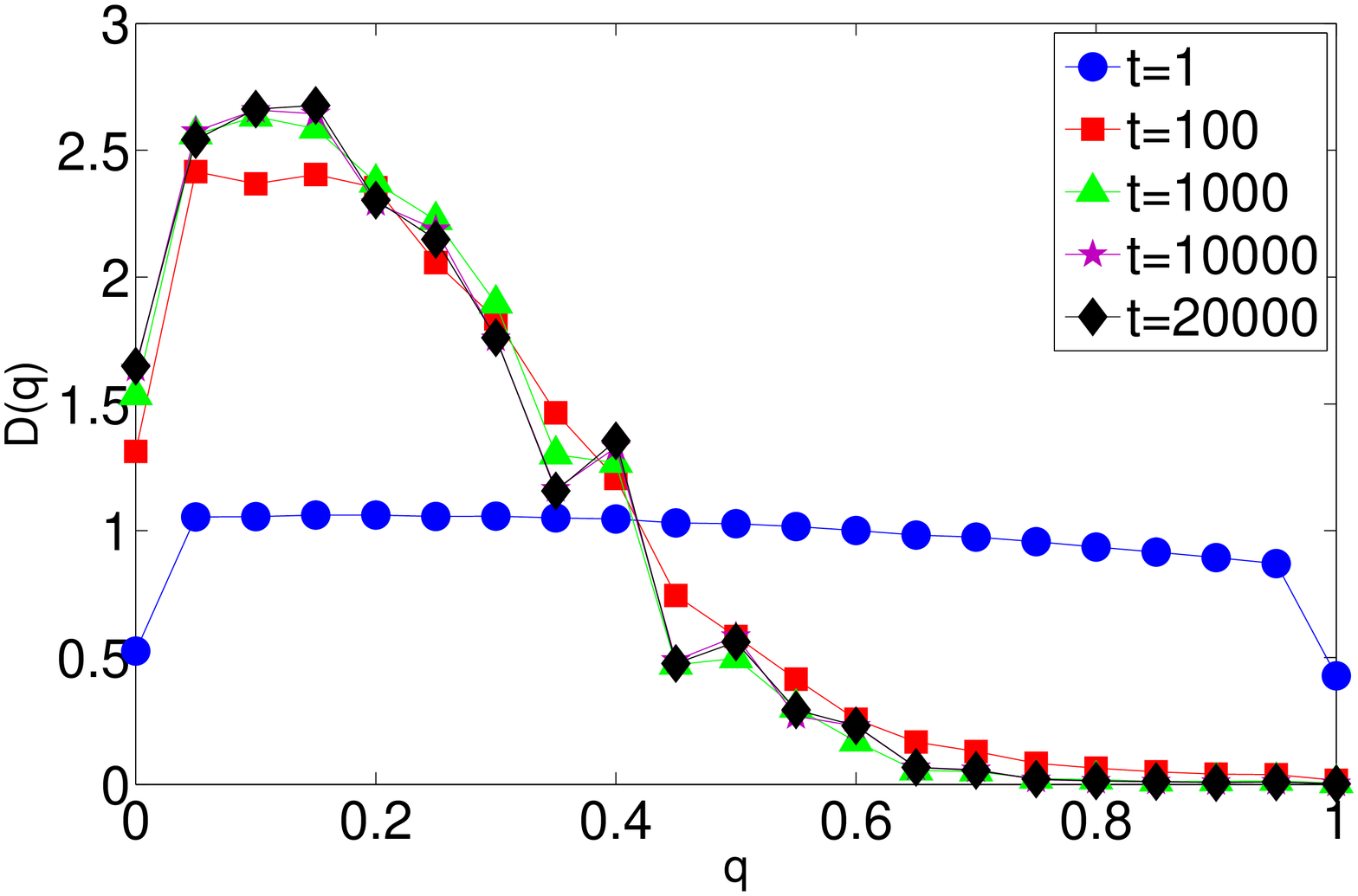}}
\caption{The distributions of offers $D(p)$ [(a) and (b)] and of
  acceptance thresholds $D(q)$ [(c) and (d)] for ER [(a) and (c)] and
  SF [(b) and (d)] networks when a Replicator-like update rule,
  eq. (\ref{eq:REP}), is at work.}
\label{fig:RD2}
 \end{figure}

\subsection{Networks of type C players (independent $p$ and $q$)}

Finally, we explore the situation according to which players are
allowed to choose their offers $p$ and acceptance thresholds $q$
independently. In Figures \ref{fig:RD2}.a and \ref{fig:RD2}.b we plot
the distribution of offers $D(p)$ for ER and SF networks
respectively. Remarkably the two distributions show a maximum around
$p\simeq 0.3$, pointing out that offers are quite poor in this third
setting. In the case of ER, nearly all the offers are concentrated
around the maximum and time evolution shows that large offers
dissapear first from the population, similarly to the case of players
A on ER networks. For SF networks $D(p)$ is remarkably broader having
nonzero values for the entire range of $p\in[0,1]$.  Therefore, only
in SF networks we observe some degree of altruistic behavior, although
the probability of finding offers with $p>0.5$ is lower than that for
$p<0.5$.

Turning the attention to the distribution of acceptance thresholds
$D(q)$ (Figures \ref{fig:RD2}.c and \ref{fig:RD2}.d) we observe that
both networks present quite similar behaviors since in both, players
accept low offers although they are still far from a fully
  rational behavior ($q=0$). In particular, for ER networks any offer
above $0.4$ will be accepted. In the case of SF networks this global
threshold is slightly larger although the probability of finding
acceptance thresholds with $q>0.5$ is extremely low. Interestingly, in
both distributions we find that the probability of finding players
with $q=0$ is nonzero.

\begin{figure}
\begin{center}
{\includegraphics[width=15.5cm]{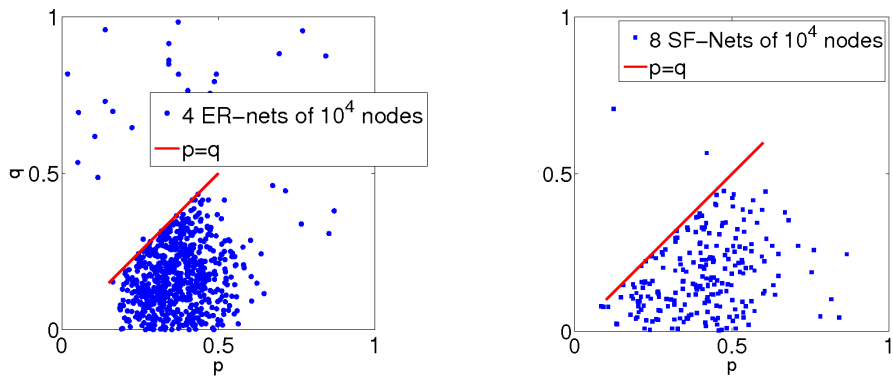}}
\end{center}
\caption{Scatter plot of the individual stragies $(p_i, q_i)$ in the
  asymtotic regime for ER (a) and SF (b) networks. For the ER case we
  have plotted the agents strategies $\{(p_i,q_i)\}$ corresponding to
  $4$ randomly chosen realizations, whereas for SF networks $8$
  realizations have been used. From the plots it is clear that in most
  cases $p_i>q_i$ in both topologies.}
\label{fig:corr-pq}
\end{figure}

We have also checked what is the correlation, if any, between the
values of $p$ and $q$ chosen by the players in order to unveil whether
there is a natural tendency towards one of the two settings A ($p=q$)
or B ($p=1-q$). In Figure \ref{fig:corr-pq} the two scatter plots are
realized by representing the set of individual strategies
$\{(p_i,q_i)\}$ observed in the asymptotic state for several
realizations of the UG dynamics. In both ER (Figure
\ref{fig:corr-pq}.a) and SF (Figure \ref{fig:corr-pq}.b) networks one
can observe that $p_i\ge q_i$ holds for most of the populations. This
tendency clearly indicates that players are neither of type A nor of
type B, although, given the low value of the average offer
$p\simeq0.3$, their behavior resemble more that of players of type A.

\subsection{Degree of Selection}
From the scatter plots in Figure \ref{fig:corr-pq} we observe that the
strategies in ER networks fill more densely the unit square than in SF.
This result points out that the selection of strategies is larger for
SF networks, {\em i.e.} the number of strategies that survive in SF
networks after Natural selection is remarkably lower than for
homogeneous networks.  

\begin{figure}[t!]
\begin{center}
\includegraphics[width=0.5\textwidth]{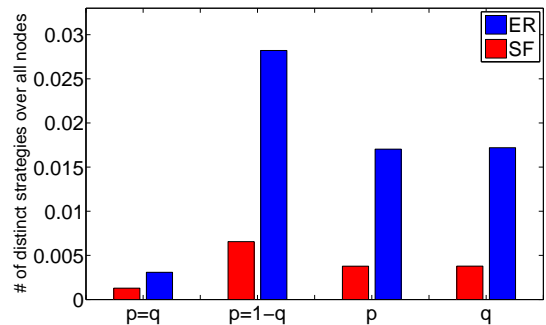}
\end{center}
\caption{Degree of selection, measured as the number of different
  asymptotic strategies divided by $N$, for ER and SF networks in the
  three different settings: (A) $p=q$, (B) $p=1-q$ and (C) $p$ and $q$
  independent.}
\label{fig:number}
\end{figure}

In Figure \ref{fig:number} we report the fraction of different
  strategies found in a population of ER and SF networks once the
  dynamical equilibrium is reached. It is clear that in SF networks
selection acts stronger than in homogeneous populations since after
selection takes place only a few number of strategies remain. We have
checked that this is due to the presence of hubs and their ability for
replicating their strategies across their surroundings (that usually
involve a large fraction of the population). In particular, for the
cases of A and B players, we have already shown that a hub can play
successfully the UG with a well-mixed population using a broad range
of $p$ values; namely, in the thermodynamic limit
($k_h\rightarrow\infty$), we have $p_h\in[0,1]$ for type A and
$p_h\in[1/2, 1]$ for type B. Any of these values of $p$, when
replicated by the well-mixed population in the next generations,
increase the payoff difference between hubs and the rest of the
individuals. Therefore, the dynamics of the well-mixed population in
contact with the hub is finally frozen with the $p$ value dictated by
it. From Figure \ref{fig:number} it becomes clear that the same
  happens for populations of C players. Note also that the fact that
  the number of different strategies observed during the equilibrium
  of SF networks is smaller than that in ER networks is not
  inconsistent with the fact that the distribution $D(p)$ in SF
  displays long tails since this distribution is constructed averaging
  over many different equilibria.

 \section{Social Penalty}
In this section, we change the scenario for the selection rule of
strategies focusing on the application of the so-called ``social
penalty'' after each round robin of the UG. Let us remark that, with
this evolutionary rule, in order to survive a player has to take care
not only of its payoff, but also of those of its neighbors, since the
poorest player of the network is replaced together with all its
neighbors. Therefore, if an individual exploits his neighborhood so
that he takes a large stake of the total reward, he would risk to be
dropped out of the game as a result of one of his neighbors being that
with the lowest payoff in the population of players. Consequently,
what drives the evolution of the distribution of $p$ values among the
population is the balance between the conflicting interests of earning
more (to avoid being the poorest) and earning less (to avoid being
stigmatized). This conflict could, in principle, be solved in the case
of hubs in SF networks: being the most connected elements, hubs are
topological favoured to accumulate a large payoff per round. Therefore
a hub can afford large degrees of altruism providing his neighbors
with enough payoff to survive and, at the same time, without any risk
of being himself the poorest element of the population.

Notice that, at variance with Natural Selection, successful strategies
do not replicate but simply survive in the long term. Therefore, as
the removed individuals are replaced by new players with randomly
chosen strategies the equilibrium is approached slower than in
networks driven by Natural Selection. The results presented below
correspond to the numerical simularions of the UG dynamics over times
up to $t=10^7$, and averaged over at least $10^2$ different
realizations of the networks and initial conditions.

%Since the update rule acts at a global level it is therefore
%difficult to argue what kind of strategies will survive in the long
%term. One can

\subsection{Networks of type A players ($p=q$)}

\begin{figure} 
 \centering
\subfigure[ER $p=q$]
   {\includegraphics[width=7cm]{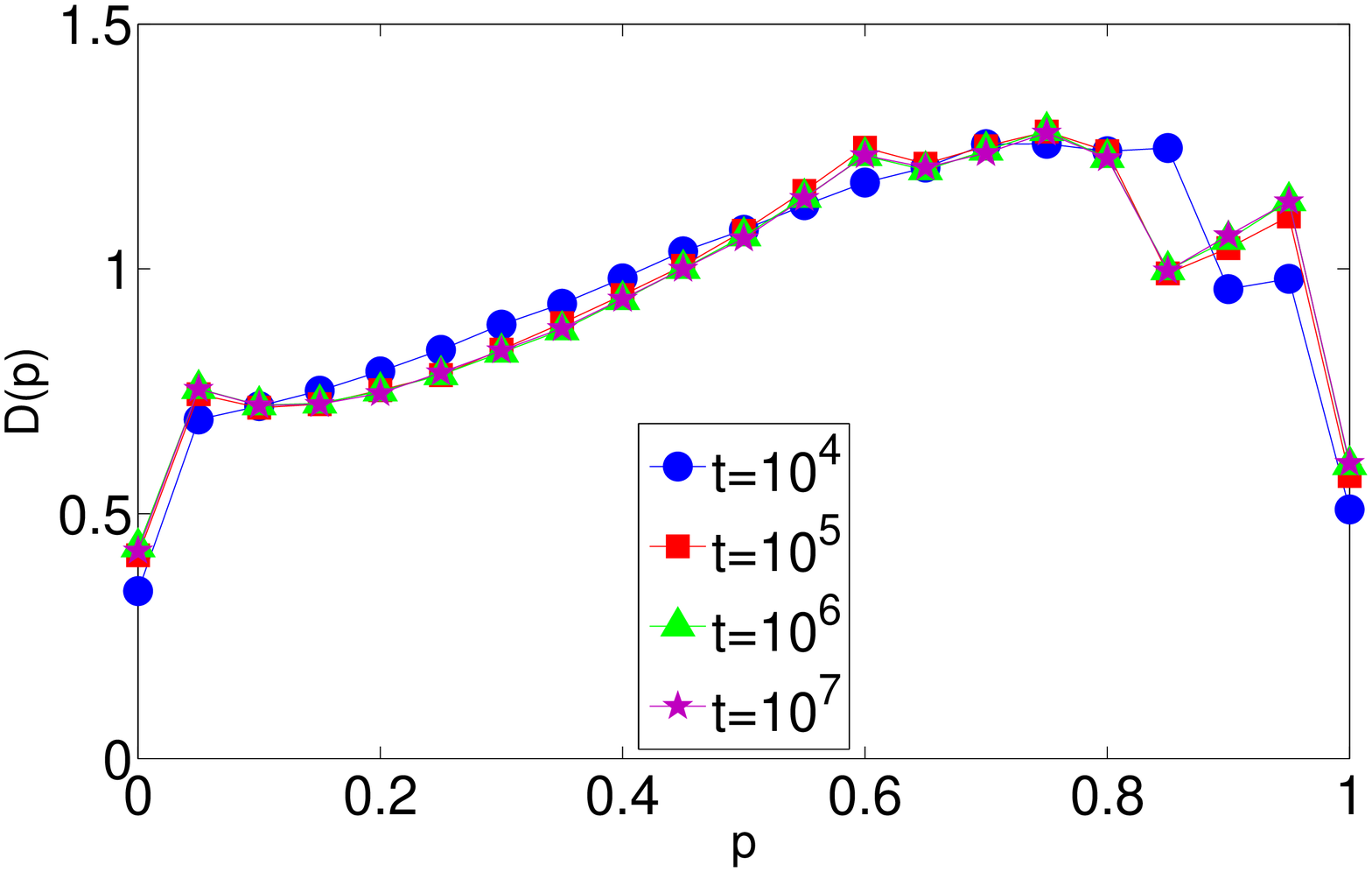}}
\subfigure[ER $p=1-q$]
   {\includegraphics[width=7cm]{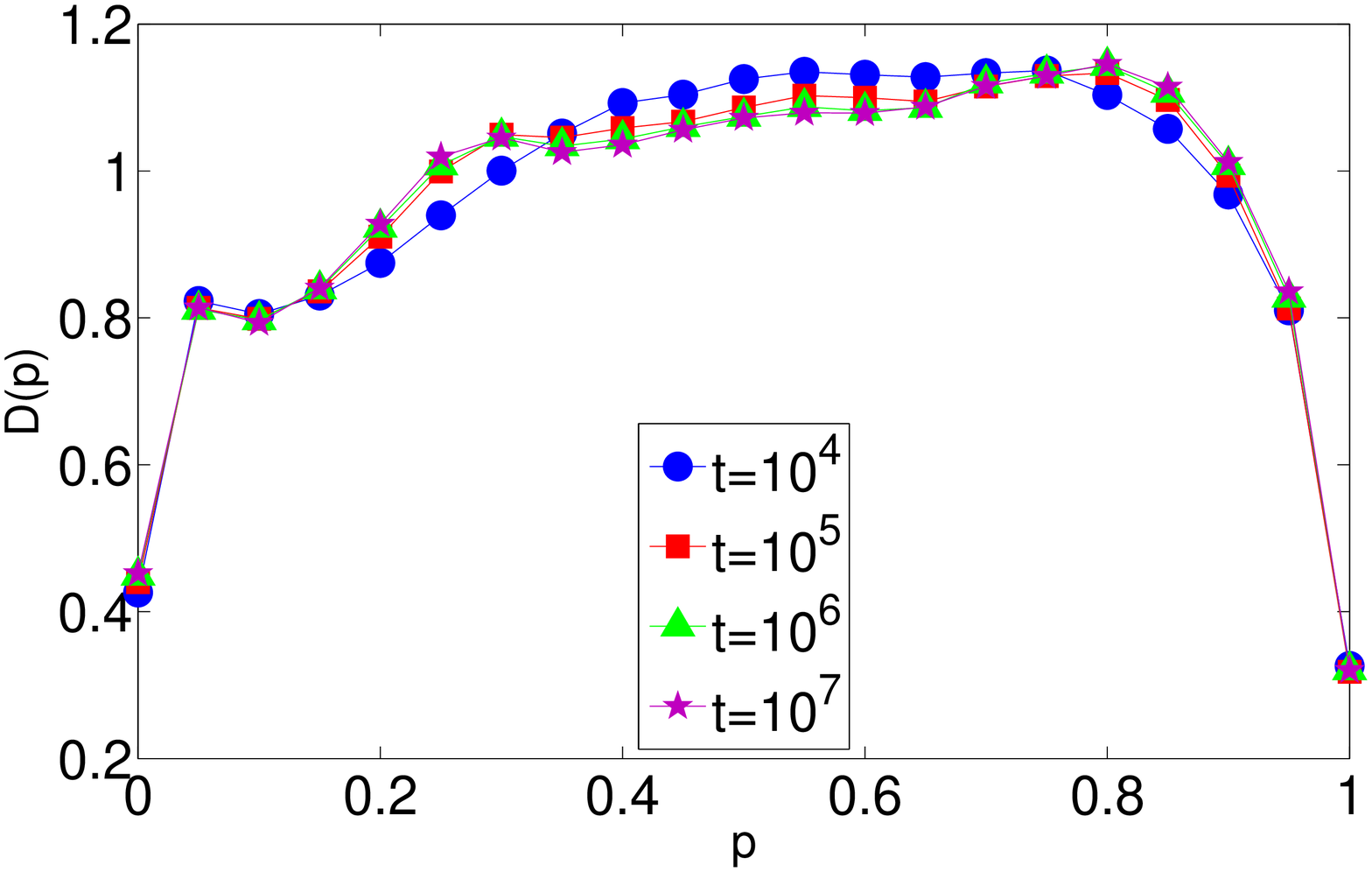}}
\subfigure[SF $p=q$]
   {\includegraphics[width=7cm]{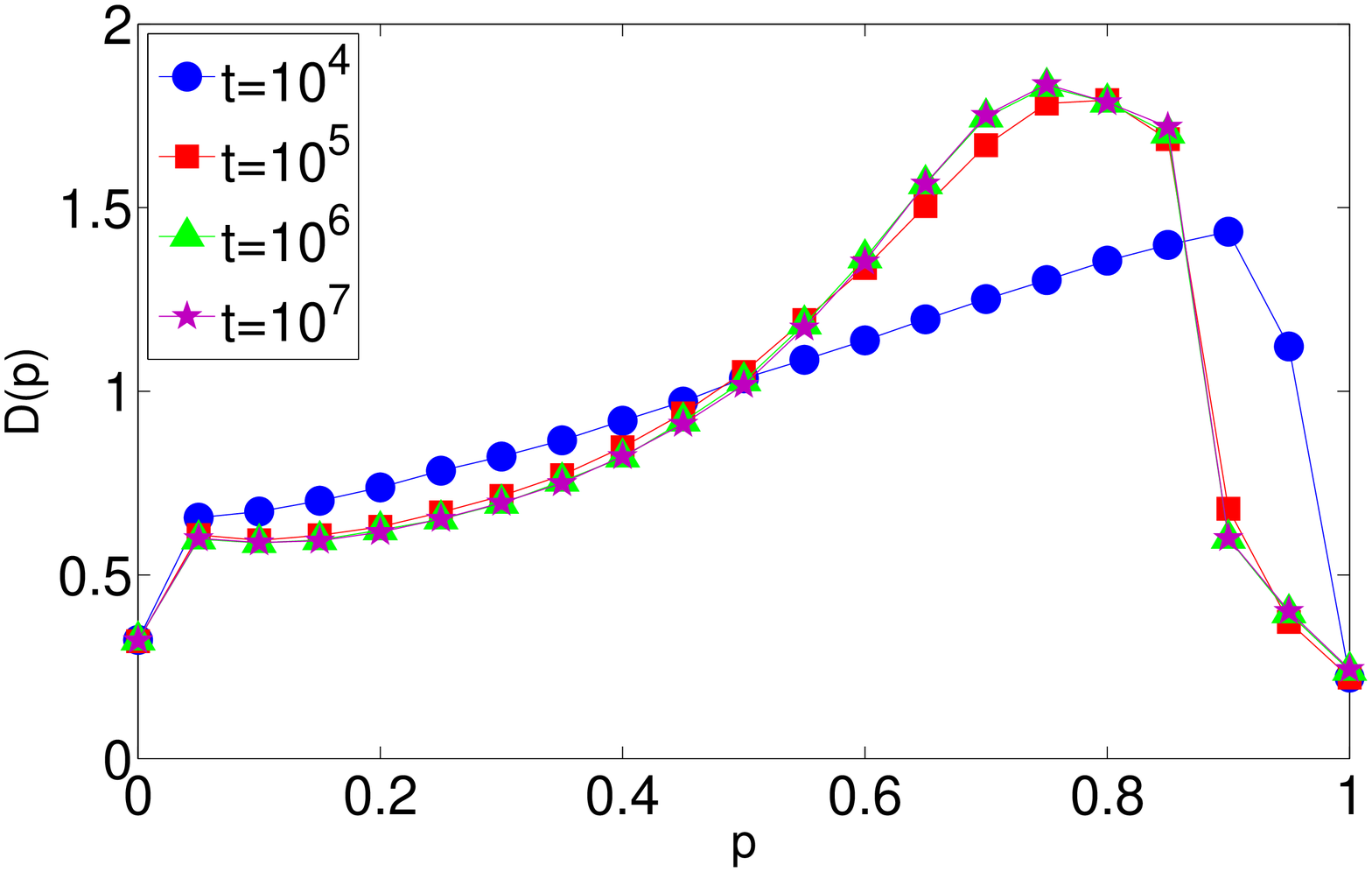}}
\subfigure[SF $p=1-q$]
   {\includegraphics[width=7cm]{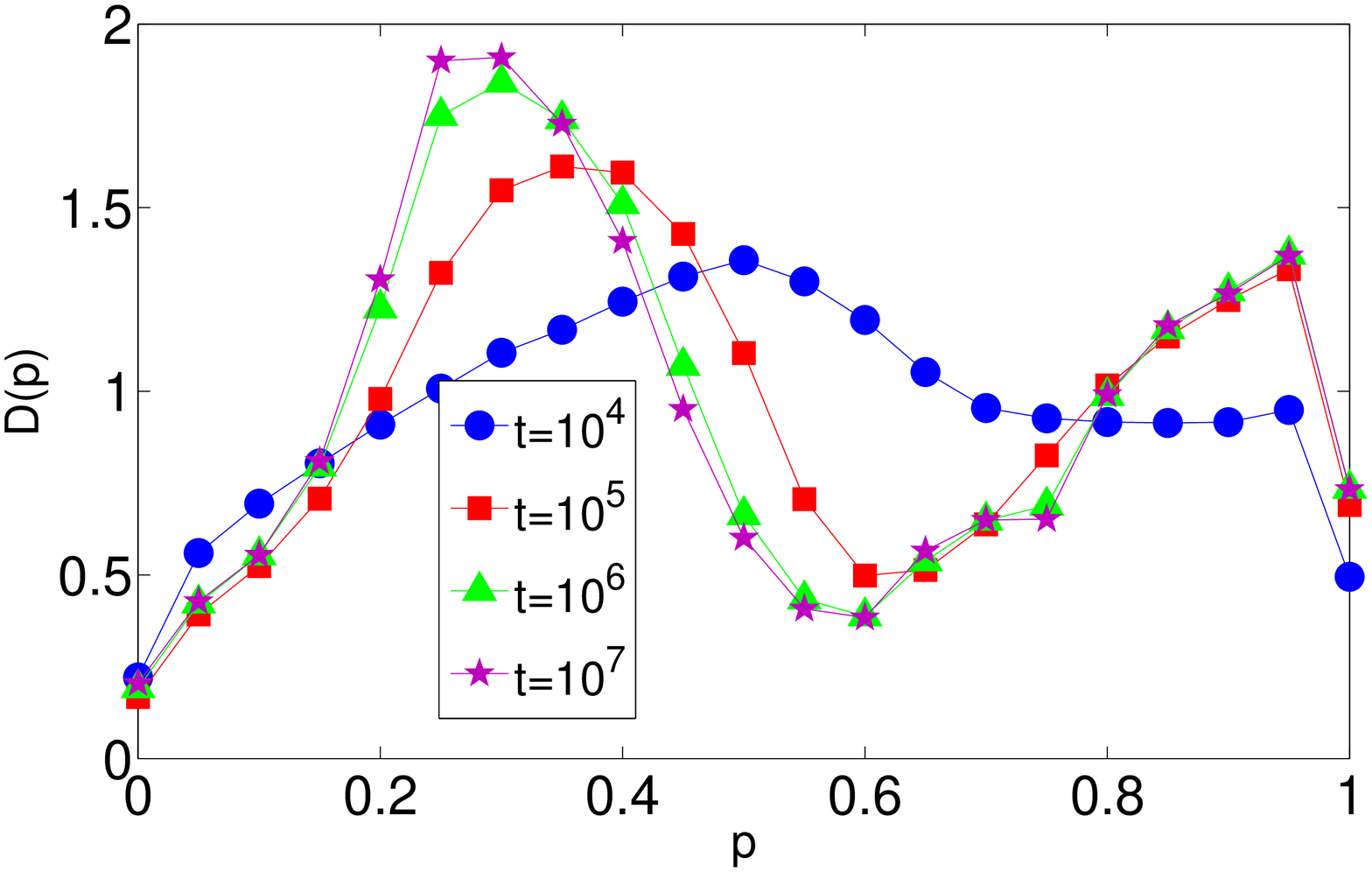}}
\subfigure[SF $p=q$]
   {\includegraphics[width=7cm]{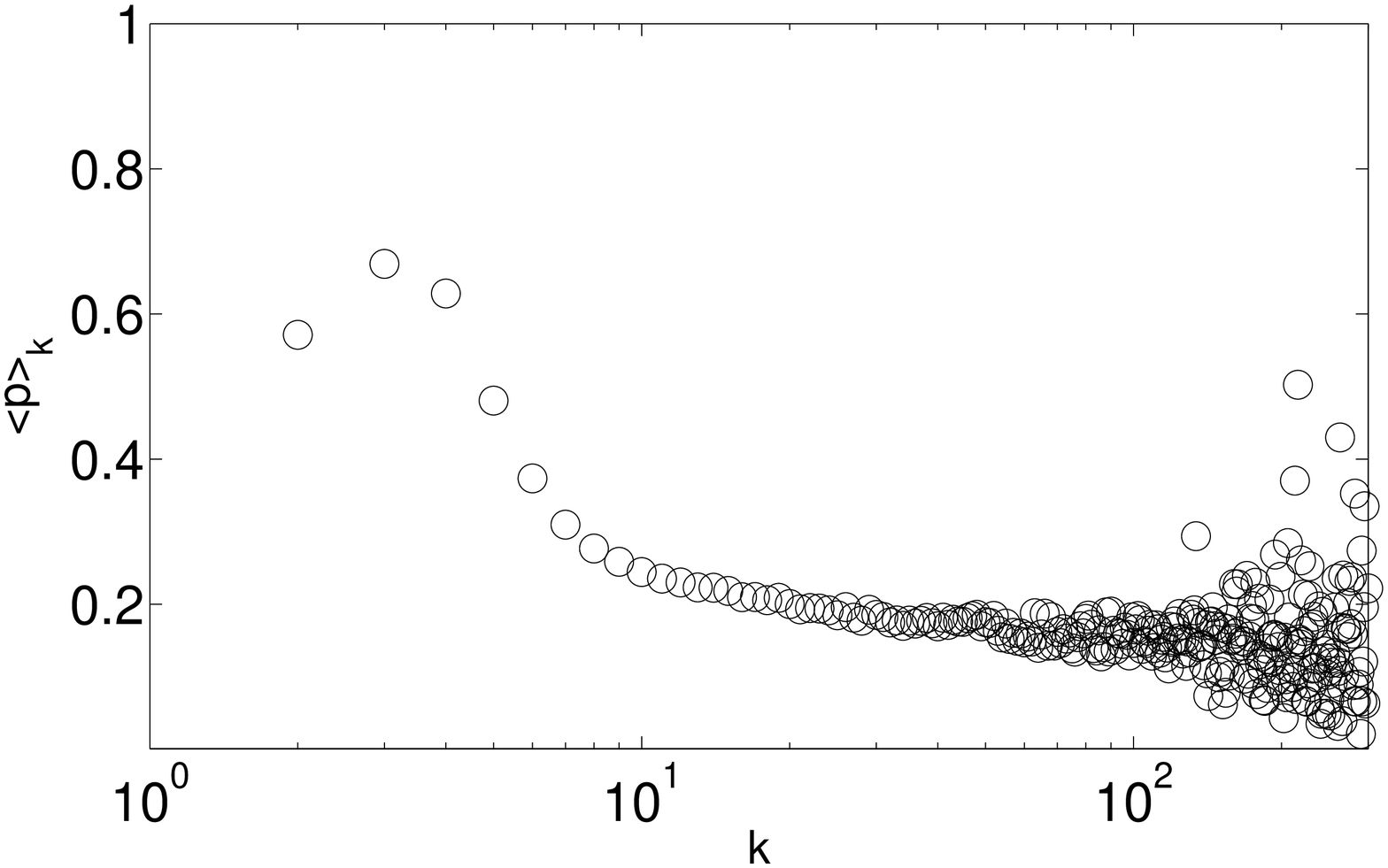}}
\subfigure[SF $p=1-q$]
   {\includegraphics[width=7cm]{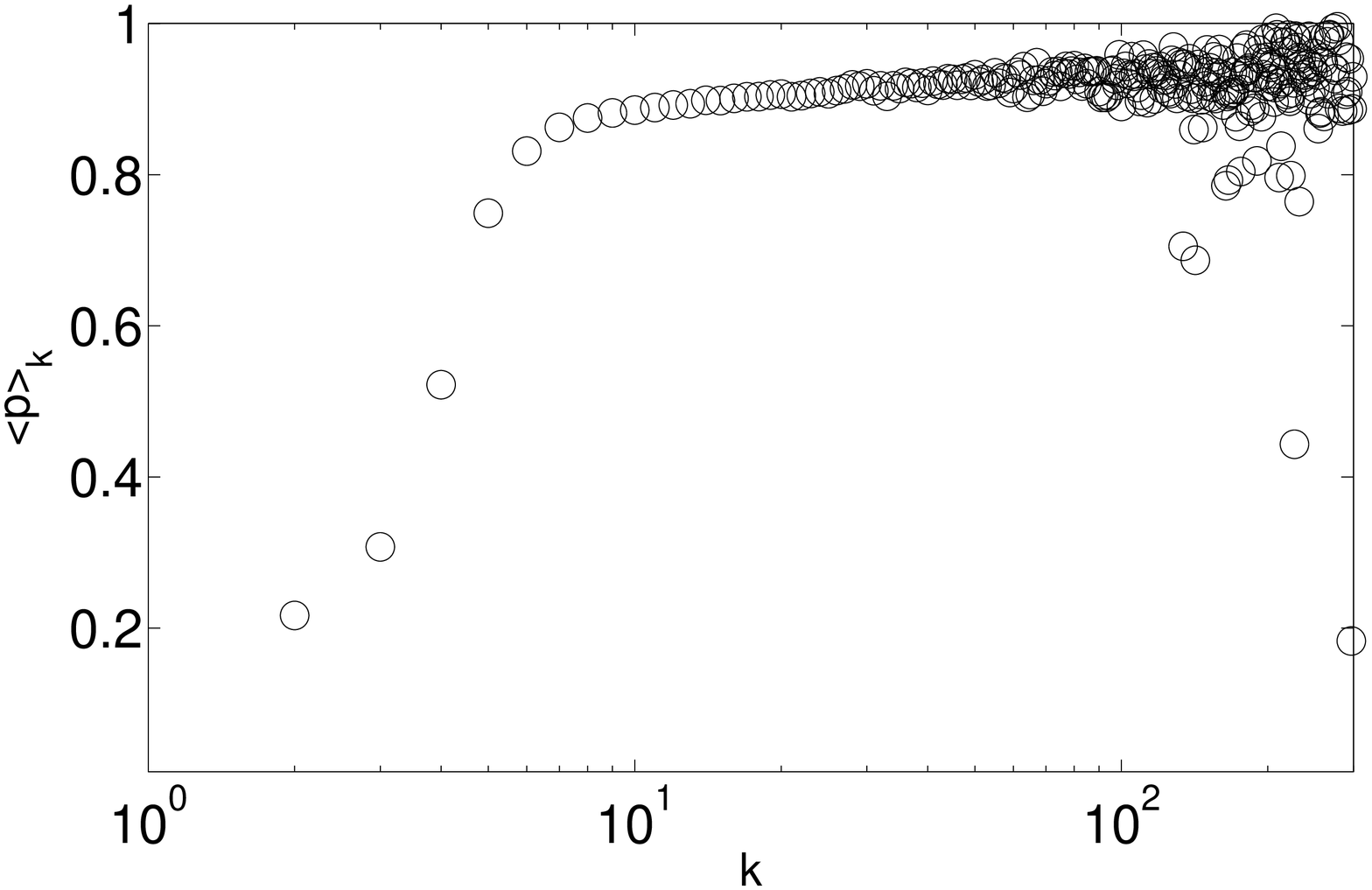}}
\caption{Distribution of offers $D(p)$ for ER and SF networks in the
  cases $p=q$ [(a) and (c)] and $p=1-q$ [(b) and (d)] when a Social
  penalty is used as the update rule. Panels (e) and (f) show the
  values of $\langle p\rangle_k$ as a function of $k$ for the cases
  $p=q$ and $p=1-q$ in SF networks.}
\label{fig:BS1}
 \end{figure}

In Figures \ref{fig:BS1}.a and \ref{fig:BS1}.c we show the evolution
of the distributions of offers $D(p)$ of type A players at different
times. In the case of ER networks (Figure \ref{fig:BS1}.a) the
distribution is nearly flat (with slowly decreasing tails at both
extremes), pointing out that any strategy can survive in a population
of type A players with homogeneous degree. On the other hand, the case
of SF networks (Figure \ref{fig:BS1}.c) reveals a more selective
population since a large number of individuals offer a quantity around
$p\simeq0.75$. However, although having a well defined maximum, it is
evident that nearly all the offers can survive.

The maximum of SF networks can be explained by looking at the mean
offer of players with degree $k$:
\begin{equation}
\langle p\rangle_k=\frac{\sum_{\{i|k_i=k\}}p_i}{NP(k)}\;.
\end{equation}
Figure \ref{fig:BS1}.e plots this quantity as a function of the degree
$k$. It is evident from the figure that those players with low
connectivity (the largest part of the population in SF networks) are
the ones playing with the offers around $p\simeq 0.75$. On the other
hand, offers from high degree nodes are very low. This latter result
points out that hubs are far from being altruistic in the case of a
population of type A players. Moreover, in the case of a hub connected
to a large number of low degree nodes, the offers from the hub will be
automatically rejected since $p_h$ is lower than those offered by the
leaves. Besides, since most of leaves offer $p>1/2$ to the hub, it
takes the largest part of the reward in all its interactions with the
leaves. Therefore, hubs exploit their neighboring leaves in a
population of type A players, thus contradicting the arguments about
the need of generosity from hubs when social penalty is at work.

\subsection{Networks of type B players ($p=1-q$)}

In the case of type B players the stationary distribution of offers
$D(p)$ for ER and SF networks are shown in Figures \ref{fig:BS1}.b and
\ref{fig:BS1}.d respectively. Interestingly, both distributions show
the same average value for the offers $\langle p\rangle\simeq
0.5$. Though of equal average value, the distribution densities are
strikingly different for both kind of networks. While for ER networks
$D(p)$ is almost flat with slowly decreasing tails at both extremes (such as in the case of type A players), it is bimodal for
the SF network. The two local maxima of $D(p)$ in SF networks are
placed at $p\simeq 0.3$ and $p\simeq 1$. In principle this result
points out the polarization of the population into altruistic and
selfish individuals. Therefore, the degree-heterogeneity of SF
networks promotes a very different microscopic balance of conflicting
aims, as reflected in the bimodal $D(p)$, with respect to the mostly
uniform density of strategies observed in near homogeneous networks
(ER).

The answer of such bimodal distribution in SF networks can be obtained
by looking at Figure \ref{fig:BS1}.f, that shows the dependence of
$\langle p\rangle_k$ on the degree of the nodes. In this case the mean
offer is seen to increase with the degree, in agreement with the
expected behavior for high degree nodes in SF networks explained
above. Moreover, the hubs of the network display a complete altruistic
behavior $p\rightarrow 1$. In this way, since the relation between the
offers of two players $p_i+p_j\geq 1$ must hold in order to conclude a
deal, low degree nodes attached to hubs both achieve the former
successful combination of offers and maximize its reward by chosing
low values of $p$.

It is possible to show that, within the context of a SF network of
type B players, hubs can afford full generosity without any risk.  Let
us define the "interacting degree" of node $i$, $k_i^{int}$, as the
number of neighbors of $i$ with whom it interacts successfully ({\em
  i.e.} those satisfying $p_j+p_i \geq 1$, the interacting
neighborhood). If we consider a hub in a SF network, $k_h\gg 1$, then
under the assumption that $p$ is distributed in its neighborhood
following the same distribution as in the whole network, we obtain:
\begin{equation}
k_h^{int} = k_h \int_{1-p_h}^{1}D(p)\; dp = k_h \left( 1 - F(1-p_h)\right) \;,
\label{kint}
\end{equation}
where $F$ is the (cumulative) distribution function of $D(p)$. Under
the same assumptions, it follows that the payoff received by a hub is
\begin{equation}
\Pi_{h} = k_{h} \left[ 1-F(1-p_{h})\right]
\left[(1-p_{h}) + \int_{1-p_{h}}^1 p \; dF \right] \;,
\label{pi_hub}
\end{equation}
where the integral is the average of $p$ in the "interacting"
neighborhood of the hub. Provided that this average is larger
than $\theta_b/k_{h}$, the limit when $p_{h}\rightarrow 1$ is
\begin{equation}
\lim_{p_{h} \rightarrow 1} \Pi_{h} > \theta_b \;.
\label{bounds}
\end{equation}
If $\theta_b$ is an upper bound of $\min_i \Pi_i$, then a hub will not
have the minimum payoff even if it offers the whole stake and accepts
any offer. One can give a simple estimate for the upper bound
$\theta_b$: For $k_{min}=2$, the less connected nodes offering $0$ and
linked to two fully generous neighbors will obtain $4$. That is, we
can assume $\theta_b \leq 4$, in the argument above. In other words,
if the average value of the hubs neighbors $p_{ave}>\theta_b/k_{h}$
(which at most is $4/k_{h}$), hubs can give away almost the whole
stake. In particular, in the thermodynamic limit where $k_{h}$
diverges, they can offer $p=1$. Therefore, {\em hubs can afford full
  generosity}. Moreover, they minimize the risk of being stigmatized
by adopting high values of $p$. In other words, they not only can
afford full generosity, but also better they do if they want their
neighbors safe. 

\subsection{Networks of type C players (independent $p$ and $q$)}
% \begin{figure}
%  \centering
% \subfigure[ER $(p,q)$]
%    {\includegraphics[width=7cm]{BS_ER_p_q_diff_times.eps}}
% \subfigure[ER $(p,q)$]
%    {\includegraphics[width=7cm]{BS_ER_q_diff_times.eps}}
% \subfigure[SF $(p,q)$]
%    {\includegraphics[width=7cm]{BS_SF_p_q_diff_times.eps}}
% \subfigure[SF $(p,q)$]
%    {\includegraphics[width=7cm]{BS_SF_q_diff_times.eps}}
% \subfigure[$(p,q)$]
%    {\includegraphics[width=7cm]{BS_p_vs_k_SF_p_q.eps}}
% \subfigure[$(p,q)$]
%    {\includegraphics[width=7cm]{BS_q_vs_k_SF_p_q.eps}}
% \caption{The distributions of offers $D(p)$ [(a) and (b)] and
%   thresholds of acceptance $D(q)$ [(c) and (d)] for ER [(a) and (c)]
%   SF [(b) and (d)] networks when Social Penalty is used as the update
%   rule.}
% \label{fig:BS2}
%  \end{figure}

% \begin{figure}
% \begin{center}
% {\includegraphics[width=12.5cm]{BSsca.eps}}
% \end{center}
% \caption{}
% \label{fig:corr-pq-BS}
% \end{figure}

\begin{figure}
 \centering
\subfigure[ER $(p,q)$]
   {\includegraphics[width=7cm]{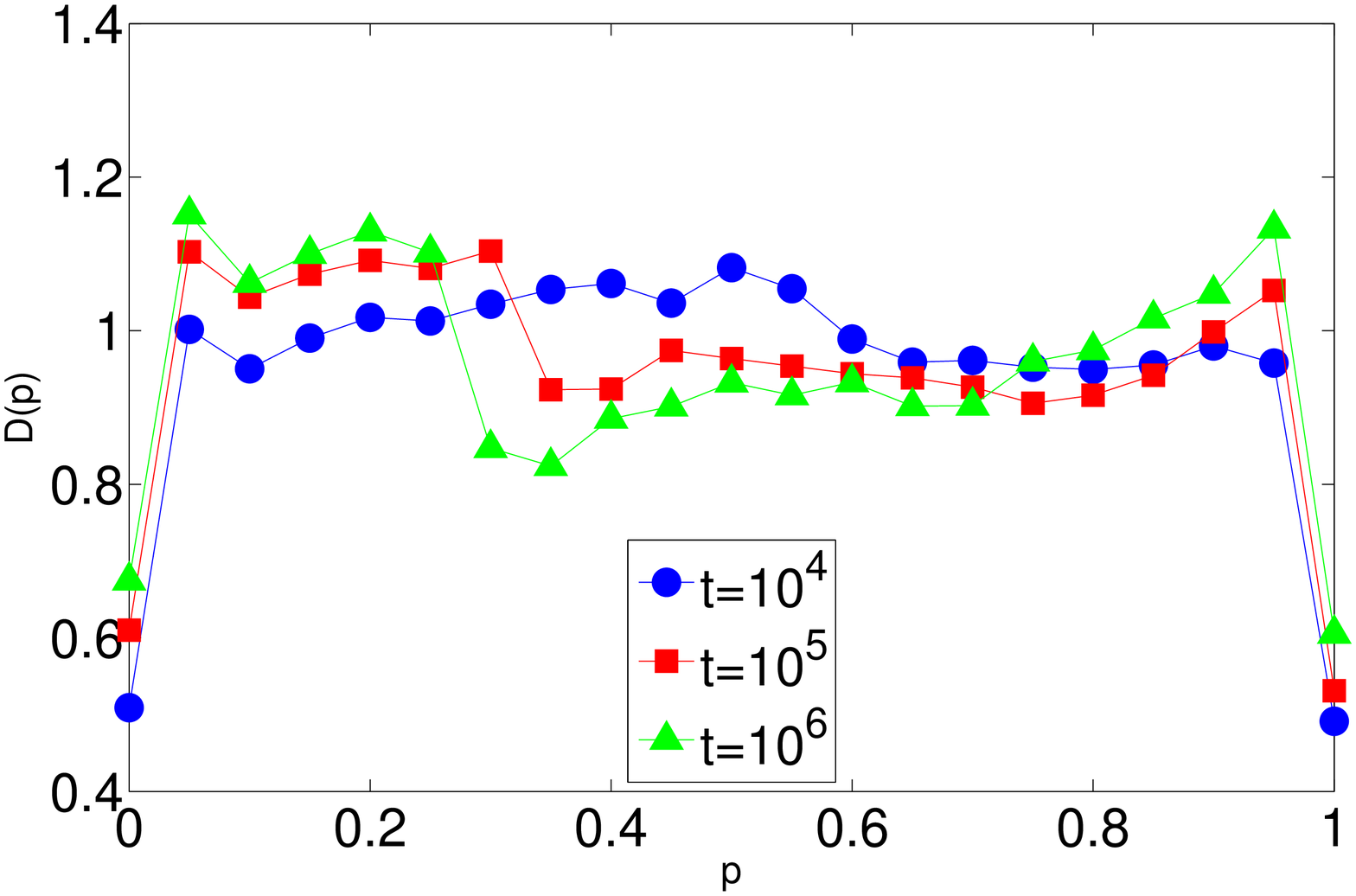}}
\subfigure[SF $(p,q)$]
   {\includegraphics[width=7cm]{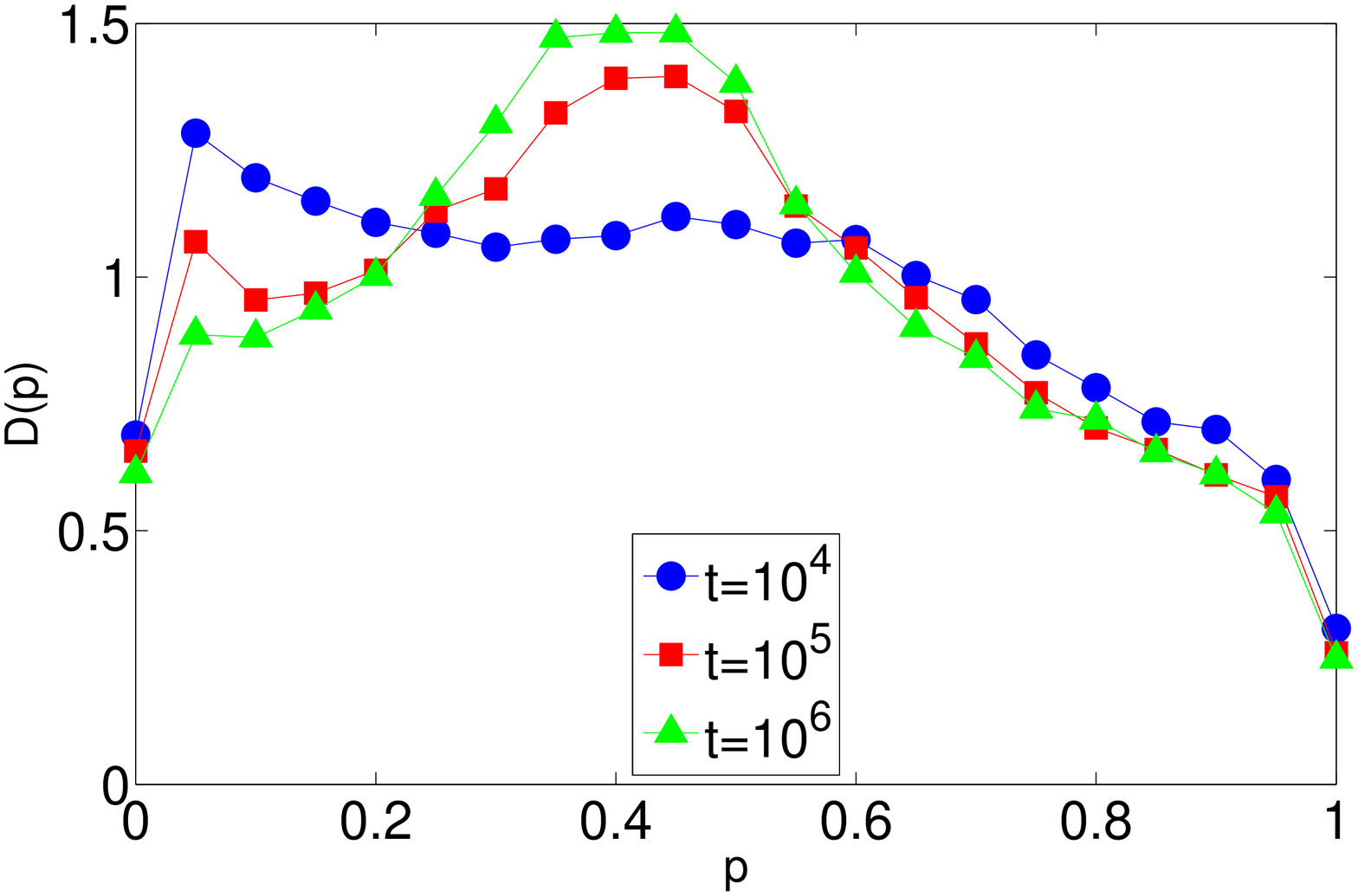}}
\subfigure[ER $(p,q)$]
   {\includegraphics[width=7cm]{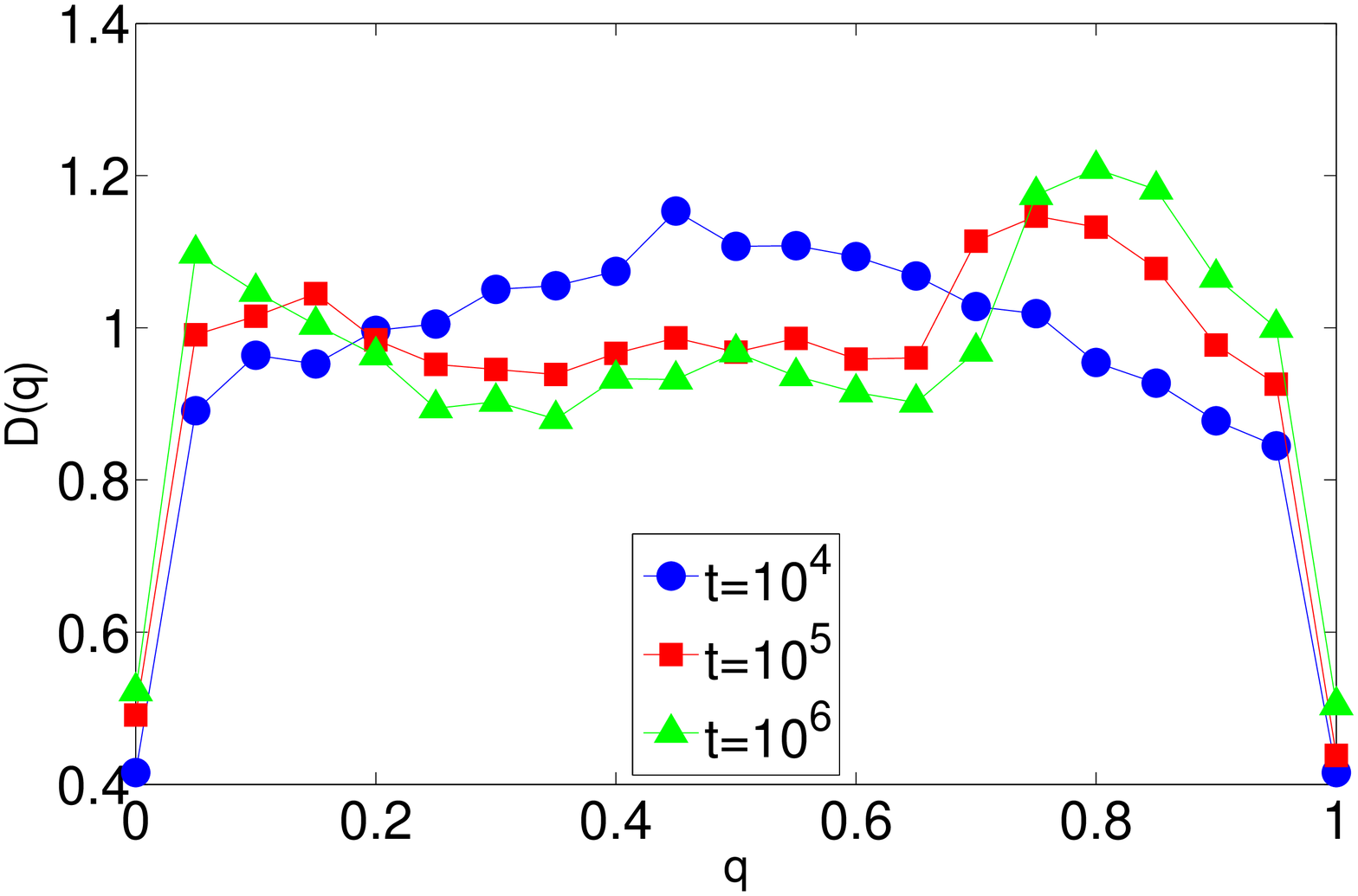}}
\subfigure[SF $(p,q)$]
   {\includegraphics[width=7cm]{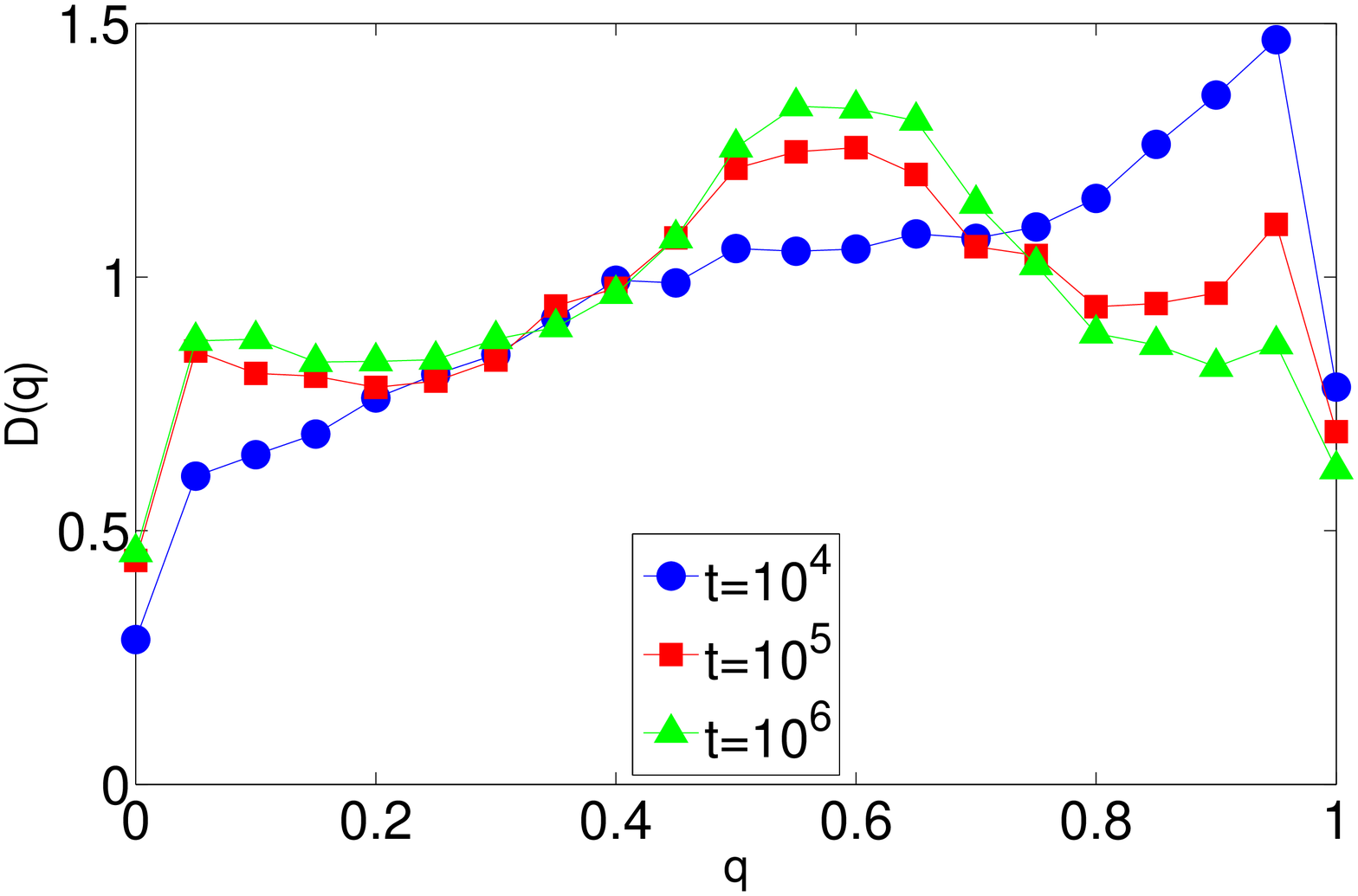}}
\subfigure[$(p,q)$]
   {\includegraphics[width=7cm]{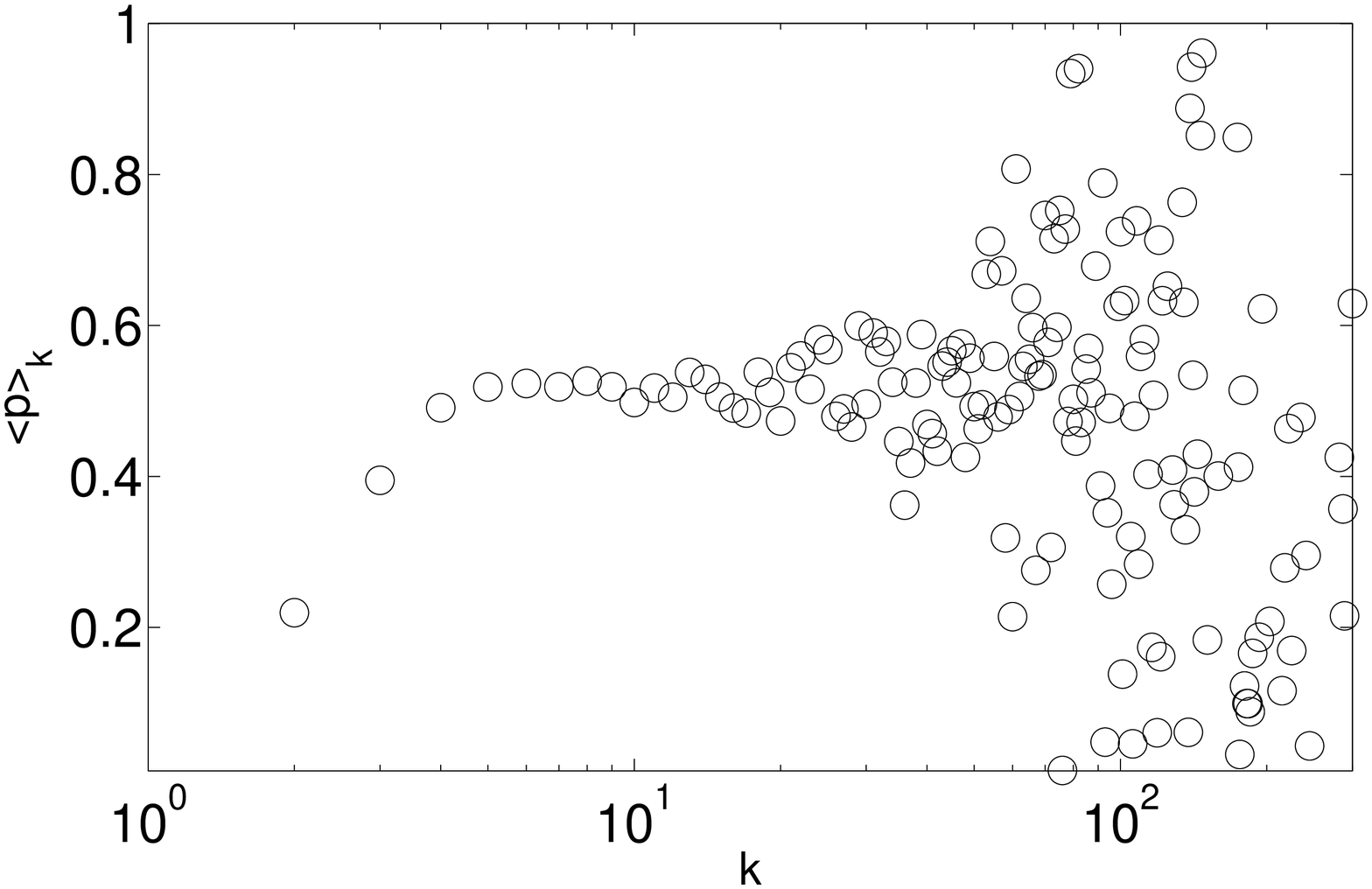}}
\subfigure[$(p,q)$]
   {\includegraphics[width=7cm]{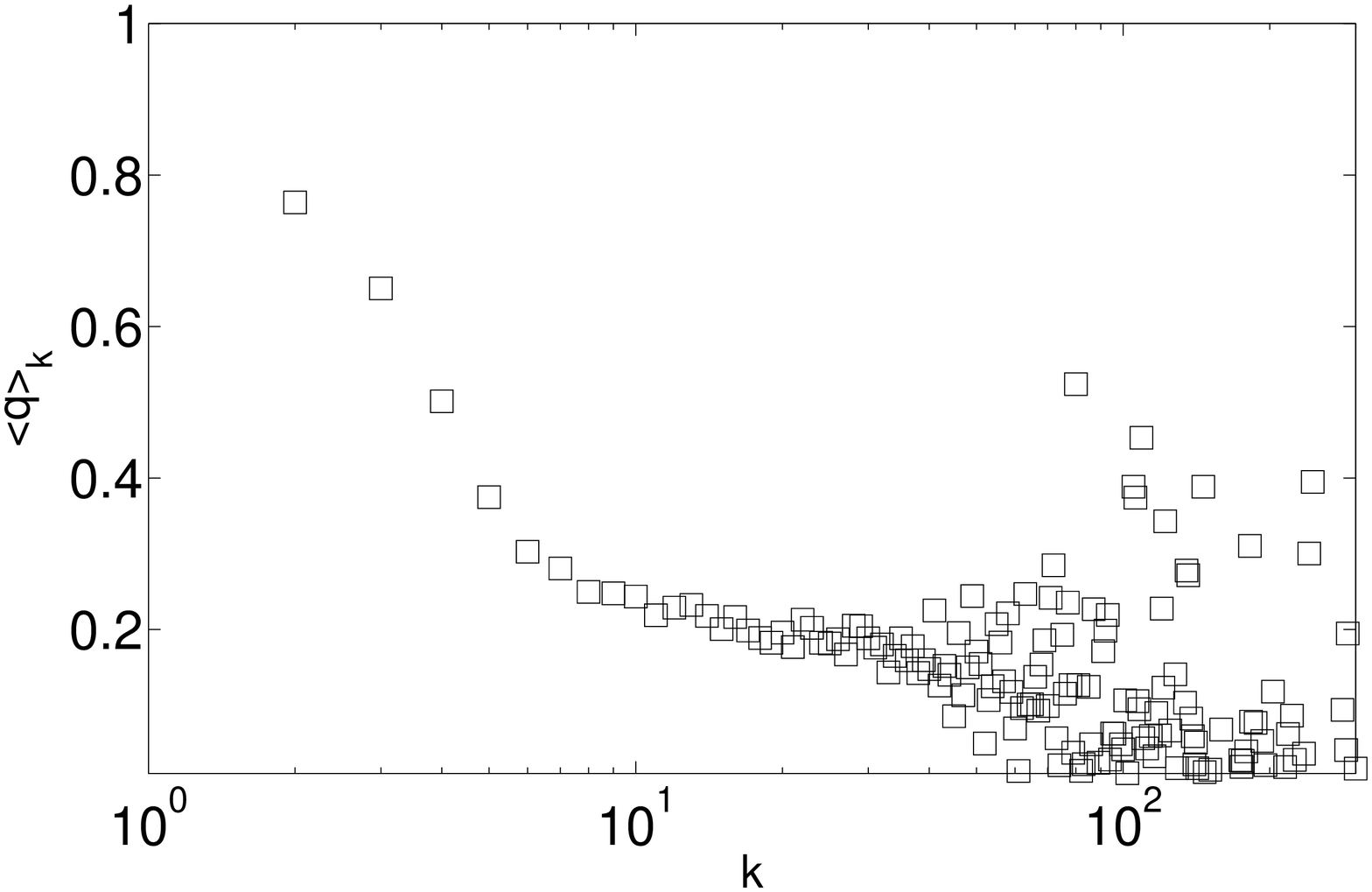}}
\caption{The distributions of offers $D(p)$ [(a) and (b)] and
  thresholds of acceptance $D(q)$ [(c) and (d)] for ER [(a) and (c)]
  SF [(b) and (d)] networks when Social Penalty is used as the update
  rule. Panels (e) and (f) show the values of $\langle p\rangle_k$ and
  $\langle q\rangle_k$ as a function of $k$ in SF networks.}
\label{fig:BS2}
 \end{figure}

We analyze now the case when the values of offers $p$ and acceptance
thresholds $q$ are independent. After having obtained quite different
results in populations of type A and type B players one of our aims
here is to unveil whether any of these latter behaviors is also
observed when players are free to decide the relation between $p$ and
$q$. In Figure \ref{fig:BS2} we sketch the main results for ER and SF
networks.

In the case of ER networks we show in Figures \ref{fig:BS2}.a and
\ref{fig:BS2}.c the time evolution of the distributions $D(p)$ and
$D(q)$ respectively. It is interesting to follow the time evolution of
both distributions. While at $t=10^4$ roughly all the offers and
acceptance thresholds are equally probable, for large enough times the
two distributions become bimodal: First, strategies having $p<0.25$
and $q>0.75$ are clearly favored, at the same time, both distributions
show a peak at low and high values of $p$ and $q$
respectively. Therefore, the two distributions are slightly polarized
towards high and low values of $p$ and $q$.

In SF netwoks the situation is completely different. In Figures
\ref{fig:BS2}.b and \ref{fig:BS2}.d we find asymptotic distributions
with a well-defined maximum at intermediate values of both $p$ and
$q$. In particular the two maxima are placed at $p\simeq0.4$ and
$q\simeq 0.6$ pointing out that population converges to an equilibrium
where the mean offer is similar to those values found in
  experiments whereas the acceptance threshold is larger than
  typically observed, pointing out an idiosincratic behavior
  \cite{henrich}. It is also interesting to report on the time
evolution of the two distributions. From the figure it is clear that
at moderate times $t=10^4$ the population focus on low offers and high
acceptance thresholds, a situation in which a few deals can be
concluded and thus the global payoff is minimum.  At $t=10^5$ the low
$p$ and high $q$ regions are abandoned and the population tends to
concentrate around the maxima of the asymptotic distributions at
$t=10^6$ and then a large amount of deals can be concluded.

Looking at the distributions of $p$ and $q$ across degree clases,
$\langle p\rangle_k$ (Figure \ref{fig:BS2}.e) and $\langle q\rangle_k$
(Figure \ref{fig:BS2}.f), we see clearly that the population occupying
the regions around the maxima of both $D(p)$ and $D(q)$ are those
players of low degree. Interestingly, in the case of $\langle
p\rangle_k$ there is a range, from intermediate to high degrees, where
a constant average offer $\langle p\rangle_k\simeq 0.5$ is reached.
Similarly, in the same range of degrees, the values of the acceptance
thresholds stabilize around $\langle q\rangle_k\simeq 0.2$. The
overall trends of both functions are that $\langle p\rangle_k$ grows
with the degree (similarly to what is found in SF networks of type B
players) whereas $\langle q\rangle_k$ decreases with $k$. This
indicates that high degree nodes, suported in their topological
advantadge, accept the low offers from the leaves and offer a large
part of the stake to them, thus favouring their survival.

\begin{figure}
\begin{center}
{\includegraphics[width=12.5cm]{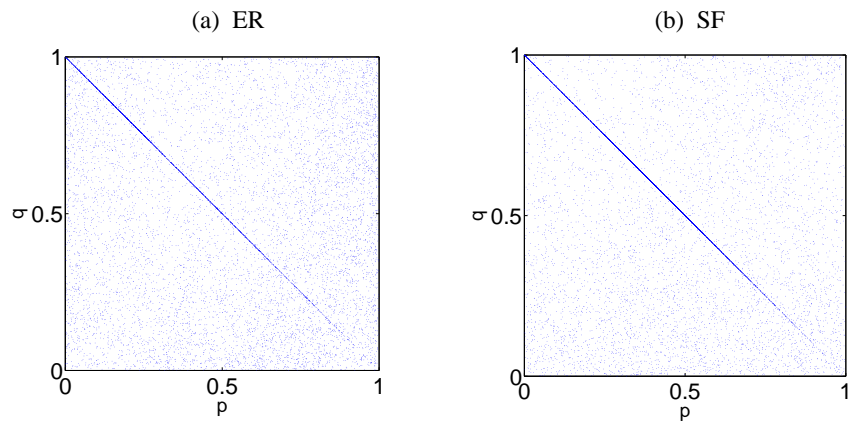}}
\end{center}
\caption{Scatter plot of the individual strategies $(p_i, q_i)$ in the
  asyntotic regime for ER (a) and SF (b) networks. For the both the ER and SF case we plot the population of $10^4$ randomly chosen realizations,  In both topologies the most frequent combination that emerges is $p_i=1-q_i$, resembling the case of players of type B.}
\label{fig:corr-pq-BS}
\end{figure}

From figures \ref{fig:BS2}.e and \ref{fig:BS2}.f we can conclude a
coarse-grained description of the population: Individuals with high
(low) values of $p$ display low (high) acceptance thresholds. Although
this description is based on average values across degree classes it
is clear that the assumption $p=q$ is no longer valid when players are
allowd to chose $p$ and $q$ freely. We have checked the true
correlation between the individuals values of $p_i$ and $q_i$ for ER
and SF networks. In Figure \ref{fig:corr-pq-BS} we show the set values
of the pairs $\{(p_i,q_i)\}$ obtained in the asymptotic
regime. Surprisingly, the accumulation of points along the curve
$p=1-q$ points out that social penalty promote the behavior as type B
players of large part of the population in both topologies.  This
result validates the assumption made above about the two strategic
groups in SF networks. Additionally, the observed trend $p=1-q$ nicely
explains the composition of the two peaks observed in the
distributions $D(p)$ and $D(q)$ in ER networks: the maximum
corresponding to large (low) offers is formed by the same individuals
that form the maximum at low (large) acceptance thresholds.

\section{Discussion and Conclusions}

We have studied the Ultimatum Game when the individulas play among
them according to a network of interactions. In the networks
considered in this study individuals can have an homogenous number of
neighbors (Erd\"os-R\`enyi graphs) or, on the contrary, present a high
degree of heterogeneity in the number of contacts (Scale-free
networks). From this perspective, we analyze how the existence of
different connectivity classes in scale-free networks affects the
behavior of the system. The Ultimatum Game dynamics has been studied
under three different frameworks: ({\em i}) role distinguishing, or
empathetic, agents (players offer the same quantity they want to be
offered), ({\em ii}) role ignoring, or pragmatic, agents (players want
to obtain the same amount both as responders and proposers) and ({\em
  iii}) agents with independent values for offers and acceptance
thresholds. Besides, we have explored two different mechanisms for
implementing the selection rule at each generation, namely: ({\em i})
\textit{Natural Selection}, according to which players replicate the
fittest agents, and ({\em ii}) \textit{Social Penalty}, according to
which, at each generation, the poorest agent is removed together with
his neighbors.

Within the context of Natural selection we have observed that the
results derived from well-mixed arguments for the case of role
distinguishing and role ignoring agents agree well with those obtained
in degree homogeneous populations, where the distributions of offers
are quite focused around $50\%$. Instead, in the case of heterogeneous
networks, the presence of highly connected nodes change quantitatively
(not qualitatively) the distribution making it broader, since hubs can
afford to make nearly all possible offers. When agents are allowed to
choose their offers and thresholds of acceptance independently, offers
tend to decrease in both Erd\"os-R\`enyi and scale-free graphs to the
$40\%$. Surprisingly, thresholds of acceptance are remarkably low, 
although they are still far from the rational economic behavior and
almost any offer above the $30\%$ of the stake is accepted. Therefore
altruistic punishment, understood as the rejection of low offers, arises in the context of Natural selection
regardless of the underlying topology.

\begin{figure}
 \centering
\subfigure[Natural Selection $(p,q)$]
   {\includegraphics[width=6.0cm]{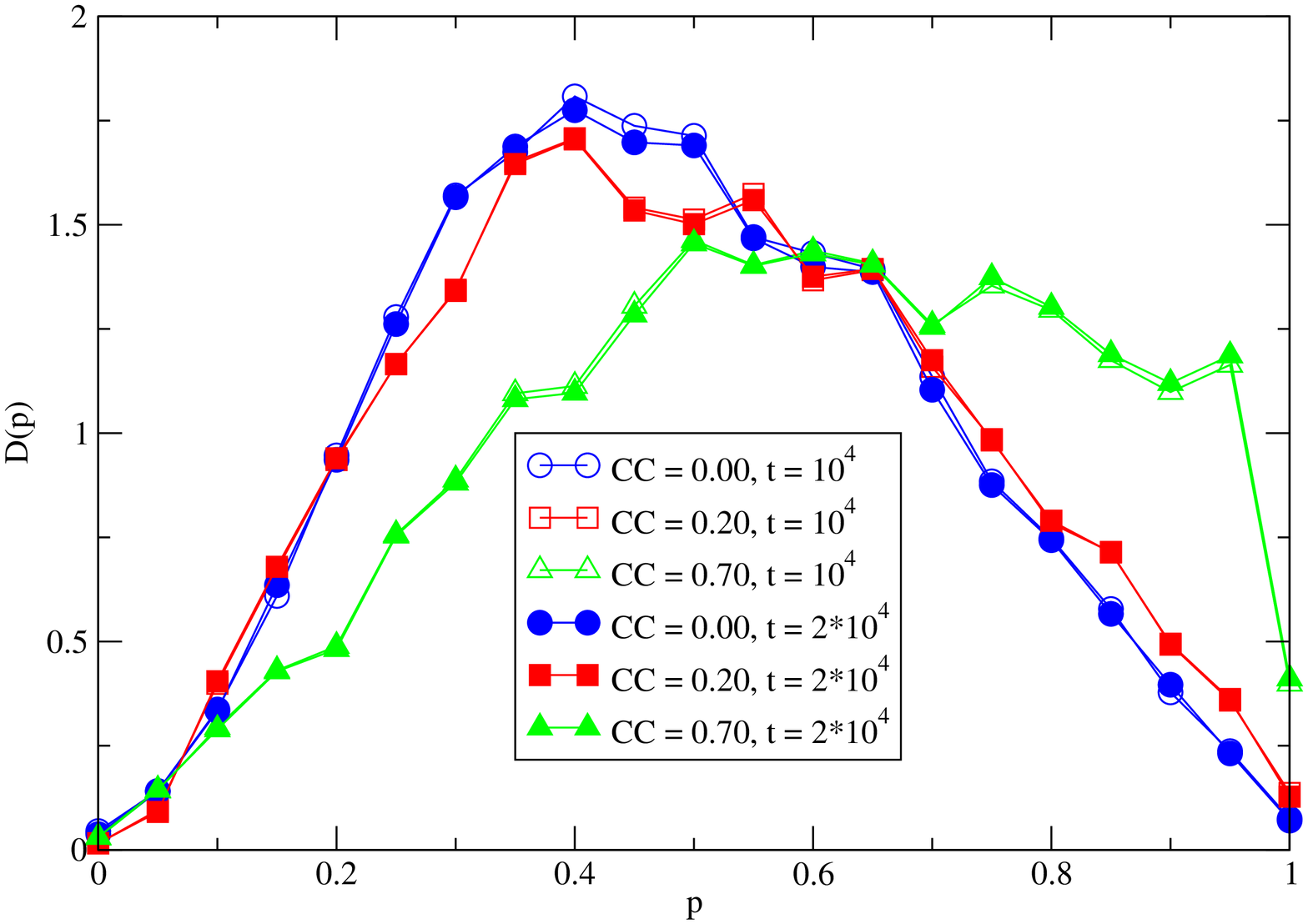}}
\subfigure[Natural Selection $(p,q)$]
   {\includegraphics[width=6.0cm]{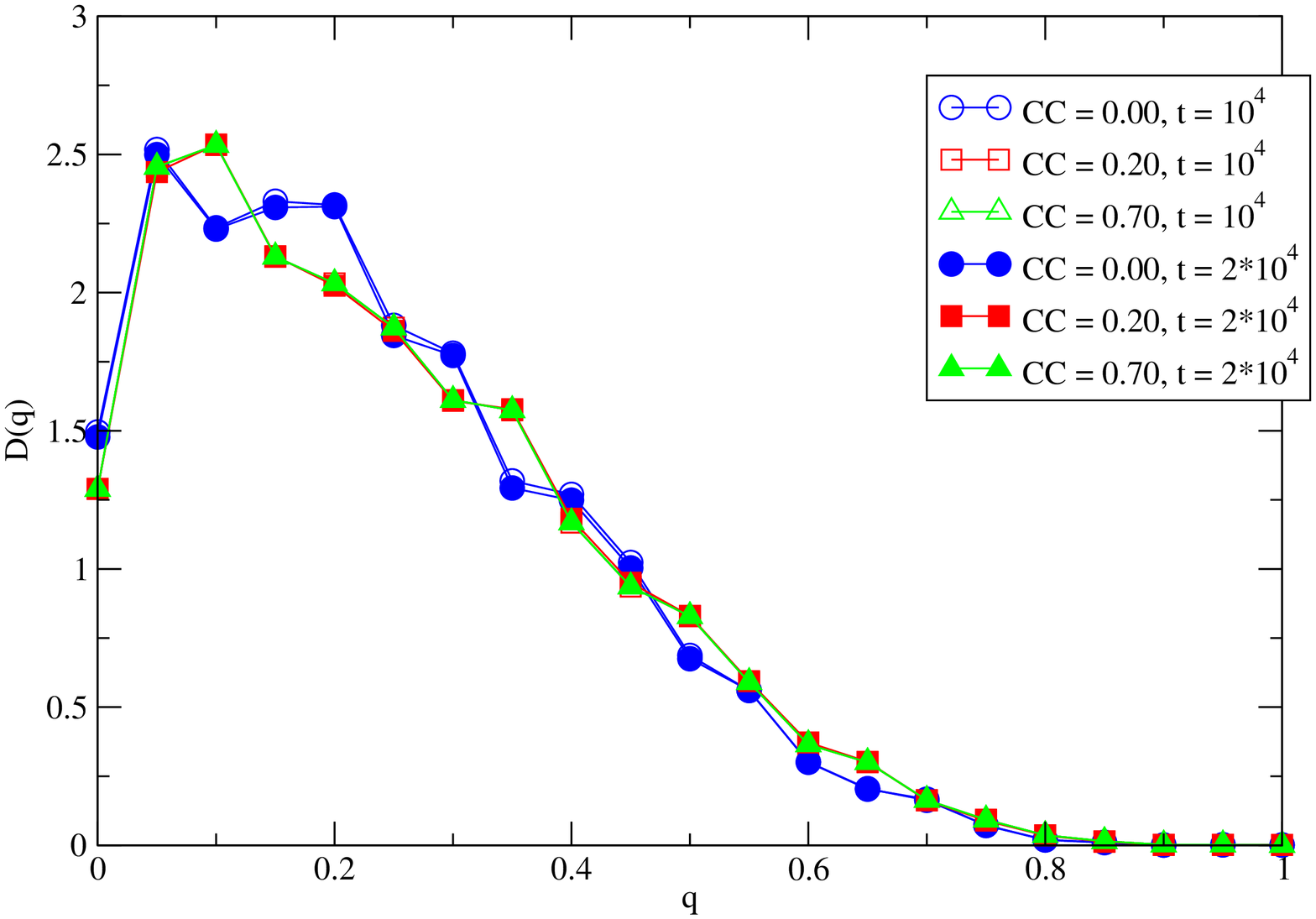}}
\subfigure[Social Penalty $(p,q)$]
   {\includegraphics[width=6.0cm]{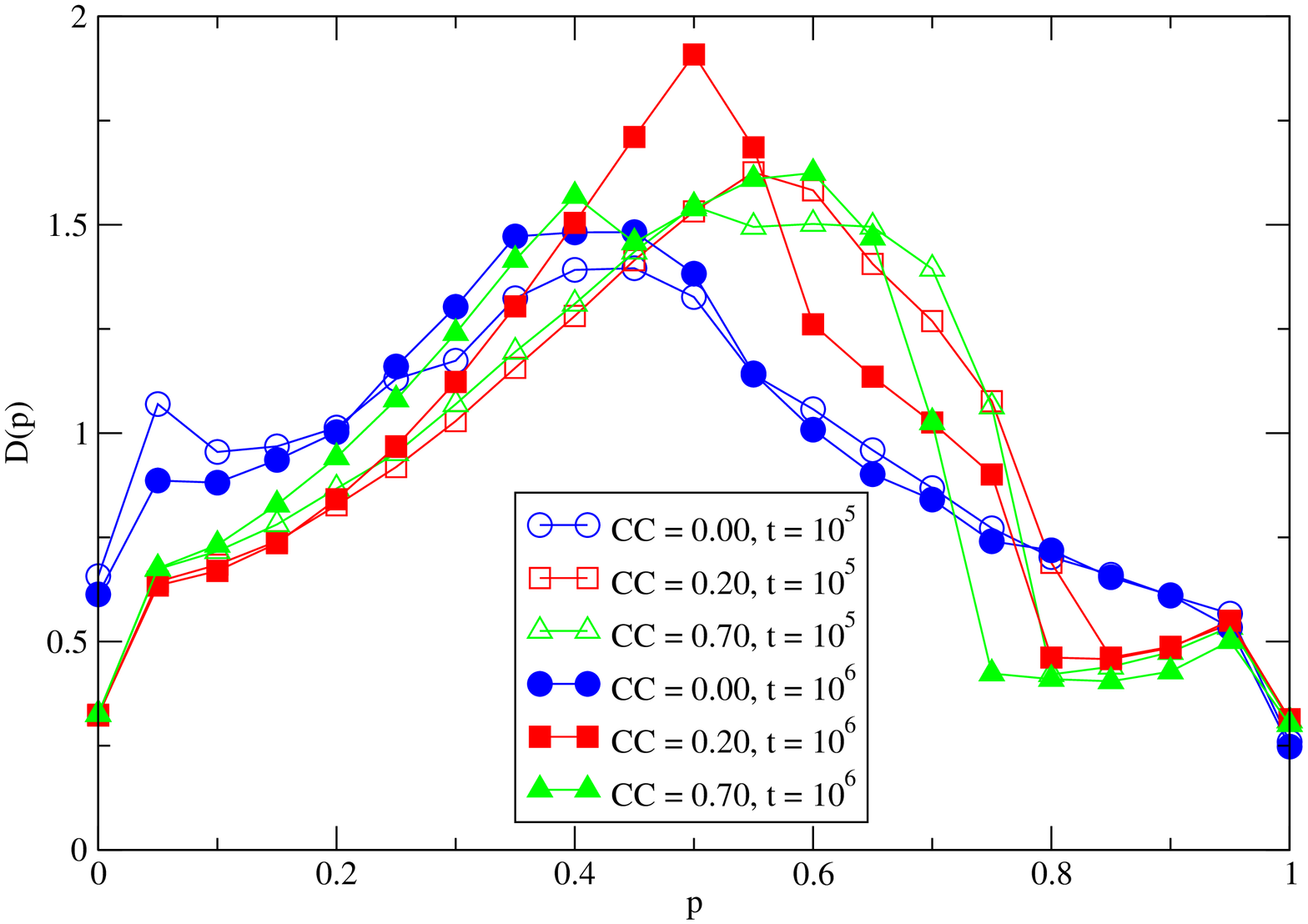}}
\subfigure[Social Penalty $(p,q)$]
   {\includegraphics[width=6.0cm]{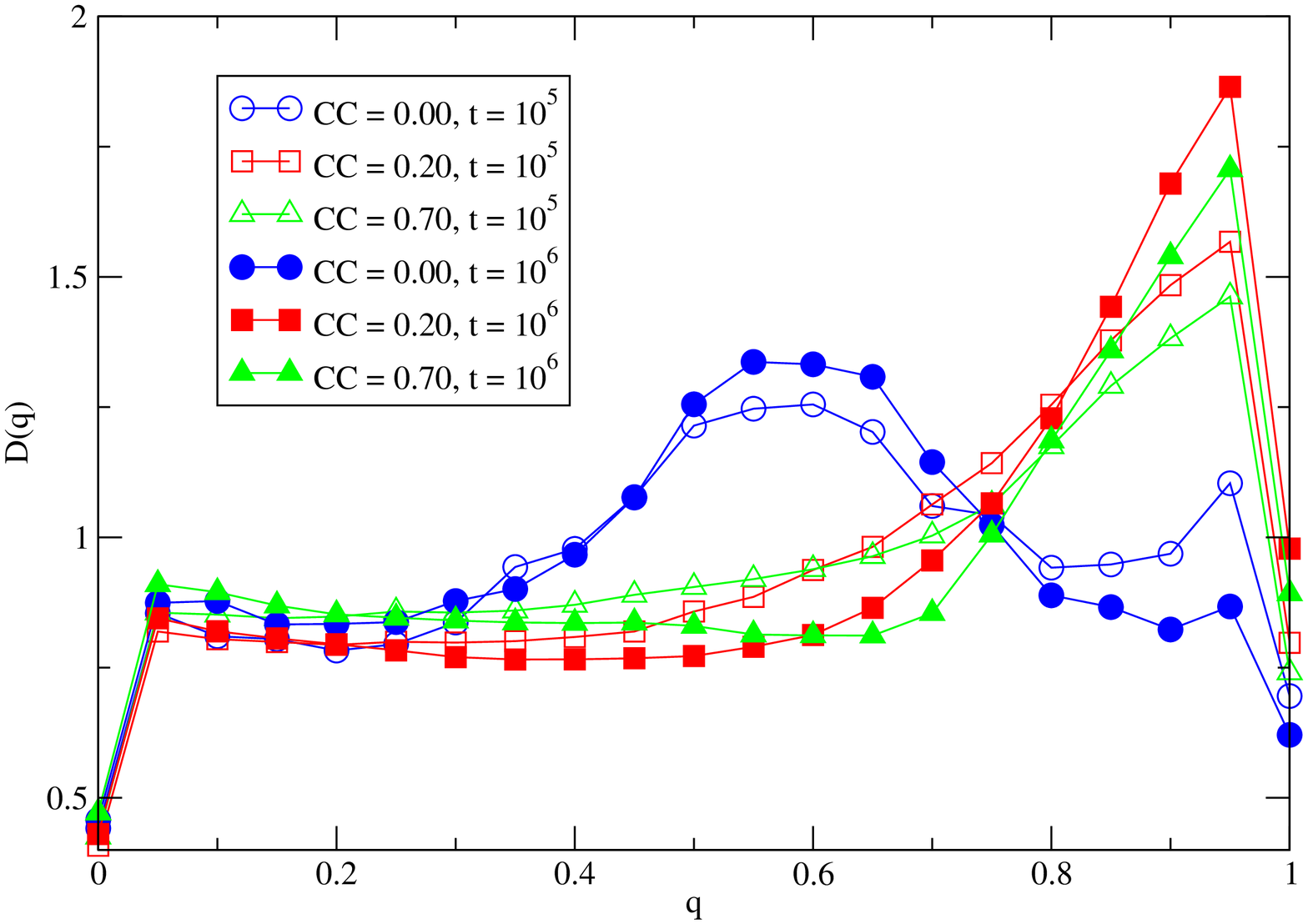}}
\caption{The distributions of offers $D(p)$ [(a) and (c)] and
  thresholds of acceptance $D(q)$ [(b) and (d)] for natural Selection [(a) and (b)]
  and Social Penalty [(c) and (d)] settings on SF networks. The networks are generated using the model in \cite{hk02} and $p$ and $q$ are independent.}
\label{fig:clust}
 \end{figure}

Interestingly, the replication of fittest strategies provokes that the
selection of strategies in the asymptotic regime is remarkably high,
especially in the case of scale-free networks. This selection is
explained in terms of the existence of hubs and their ability to
obtain a large reward with a broad range of strategies and thus to
dictate the final behavior of the entire population.

When Social punishment is implemented the dynamical behavior of the
system changes radically. Within this selection rule agents have to
take care not only about their own benefit but also about the fitness
of their neighbors. Within this context, we have found two drastically
different behaviors between empathetic and pragmatic agents. In
particular, for scale-free networks, low degree nodes and high degree
nodes display opposite behaviors in the two settings. On one hand, in
a population of role distinguishing agents, leaves are those proposing
a large portion of the stake (above $50\%$) whereas hubs show low
offers (below $20\%$). On the other hand, for role ignoring agents the
situation is the opposite, since large offers (nearly the $100\%$)
come from hubs while leaves display selfish behavior. It is therefore
in this latter setting where true altruistic behavior is
observed. Note that altruism arises in a self-organized manner with
selection acting locally: highly connected agents optimize their
chances to survive by increasing their generosity, without risking to
be the poorest in town.

Probably the most interesting result is obtained when, in the
framework of Social punishment, players can adapt their offers and
acceptance thresholds independently. Surprisingly, the dynamical
equilibria of both homogeneous and heterogeneous networks resemble to
a large extent that of role ignoring agents. In particular we have
shown that, in SF network, the large degree nodes, although not
displaying full altruism, offer a large reward (more than $50\%$) to
their neighbors and accept low offers (below $20\%$). On the other
hand, the opposite behavior is found in lowly connected players. We
have further confirmed that, in the long run, players adapt their
strategies and converge to the setting of role ignoring agents, the
framework where full altruistic behavior is observed. Let us remark
that the abundance of highly generous individuals observed when Social
Penalty is at work does not arise due to reputation \cite{nps00}, nor
costly individuals' punishment \cite{drfn08}, but from a purely 
scale-free effect combined with a social enforcement of altruism.

Finally, we point out that a full and satisfactory understanding of the models exposed here may likely demand to study the
dependence on other important topological features (such as the
clustering coefficient, degree-degree correlations, etc) or to incorporate
the competition between different kinds of individuals (role-ignoring
and role-distinguishing) into the model formulation. In particular, we have explored how our results change when the underlying SF networks have a non-vanishing clustering coefficient when $p$ and $q$ are independent (type C players). This is not an easy issue, as one should first construct networks with a tunable clustering coefficient while keeping the rest of topological properties unaltered. The model proposed in \cite{hk02} can be used to such an study as it generates scale-free networks with varying clustering properties but leaving the rest of topological features roughly the same. Our results indicate that no general conclusion can be reached as the effects of the clustering depend on several factors, of both topological and dynamical nature. As shown in Fig.\ \ref{fig:clust}, in the case of natural selection, the distribution $D(q)$ does not change when the clustering coefficient of the networks is increased from 0 to 0.7, while $D(p)$ changes if the clustering coefficient exceeds 0.2 in such a way that the average offer increases. On the contrary, for the social penalty setting, $D(p)$ remains roughly unaltered whatever the clustering of the network is, whereas $D(q)$ deviates from its behavior for non-clustered networks as soon as the clustering coefficient is increased leading to a distribution with a peak at very high acceptance thresholds. All these are aspects to further explore in future works.

\ack Y.M. is supported by MICINN (Spain) through the Ram\'on y Cajal
Program. This work has been partially supported by MICINN through
Grants FIS2006-12781-C02-01 and FIS2008-01240, and by a D.G.A. grant.

\end{document}